\documentclass[iop]{emulateapj}

\usepackage{graphicx}
\usepackage[caption=false]{subfig}
\usepackage{dcolumn}
\usepackage{bm}
\usepackage[mathlines]{lineno}
\usepackage{enumitem}
\usepackage{rotating}
\usepackage{epsfig}
\usepackage{datetime}
\usepackage{wrapfig}
\usepackage{threeparttable}
\usepackage{color}
\usepackage{soul}
\usepackage{amssymb}
\usepackage{amsmath}
\usepackage{textcomp}
\usepackage[title]{appendix}

\usepackage{natbib}

\def\simgt{\lower.5ex\hbox{$\; \buildrel > \over \sim \;$}}
\def\simlt{\lower.5ex\hbox{$\; \buildrel < \over \sim \;$}}

\def\amin{\ifmmode^{\prime}\else$^{\prime}$\fi}
\def\asec{\ifmmode^{\prime\prime}\else$^{\prime\prime}$\fi}

\def\simgt{\lower.5ex\hbox{$\; \buildrel > \over \sim \;$}}
\def\simlt{\lower.5ex\hbox{$\; \buildrel < \over \sim \;$}}

\newcommand\chandra{{\it Chandra}}
\newcommand\xmm{{\it XMM-Newton}}
\newcommand\suzaku{{\it Suzaku}}
\newcommand\integral{{\it INTEGRAL}}
\newcommand\rxte{{\it RXTE}}

\newcommand\nustar{\hbox{\it NuSTAR\/}}

\def\arcmin{\hbox{$^\prime$}}
\def\arcsec{\hbox{$^{\prime\prime}$}}

\begin{document}

\title{The Galactic bulge diffuse emission in broad-band X-rays with NuSTAR}

\author{Kerstin Perez}
\affil{Department of Physics, Massachusetts Institute of Technology, Cambridge, MA 02139, USA}
\email{kmperez@mit.edu}

\author{Roman Krivonos}
\affil{Space Research Institute of the Russian Academy of Sciences (IKI) Moscow, Russia, 117997}
\email{krivonos@iki.rssi.ru}

\and

\author{Daniel R. Wik}
\affil{Department of Physics and Astronomy, University of Utah, Salt Lake City, UT 84112, USA}
\email{wik@astro.utah.edu}

\begin{abstract}
The diffuse hard X-ray emission that fills the Galactic center, bulge, and ridge is believed to arise from unresolved populations of X-ray binary systems. 
However, the identity of the dominant class of accreting objects in each region remains unclear. 
Recent studies of Fe line properties and the low-energy ($<10$\,keV) X-ray continuum of the bulge 
indicate a major population fraction of non-magnetic cataclysmic variables (CVs), in particular quiescent dwarf novae. 
This is in contrast to previous high-energy ($>10$\,keV) X-ray measurements of the bulge and ridge, which indicate a dominant population of magnetic CVs, in particular intermediate polars.
In addition, \nustar\ broad-band measurements have uncovered a much heavier intermediate polar population in the central $\sim100$\,pc than previously assumed, raising the possibility that some fraction of this population extends further from the center.
Here we use \nustar's large aperture for unfocused photons and its broad-band X-ray range to probe the diffuse continuum of the inner $\sim$1--3$^\circ$ of the Galactic bulge.
This allows us to constrain possible multi-temperature components of the spectrum, such as could indicate a mixture of soft and hard populations. 
Our emissivity is consistent with previous hard X-ray measurements in the bulge and ridge, with the diffuse X-ray luminosity tracing the stellar mass.
The spectrum is well-described by a single-temperature thermal plasma with $kT \approx 8$\,keV, with no significant emission above 20\,keV. 
This supports that the bulge is dominated by quiescent dwarf novae; we find no evidence of a significant intermediate polar population in the hard X-ray band.
\end{abstract}


\section{\label{sec:intro} Introduction}

The Galactic diffuse X-ray emission (GDXE), an unresolved X-ray emission that fills the Galactic center and extends over $100^\circ$ along the Galactic plane, has been extensively studied since its discovery over 30 years ago~\citep[e.g.,][]{Worrall1982, Revnivtsev2006, Revnivtsev2009}.
This emission has been observed from 0.5\,keV to $>$50\,keV~\citep{Muno2004, Heard2013A, Yuasa2012, Krivonos2007}, with strong Fe emission features at 6.4\,keV, from neutral or weakly ionized Fe, and at 6.7\,keV and 7.0\,keV, from He-like K$\alpha$ and H-like Ly$\alpha$ ions~\citep{Koyama1989,Koyama1996,Xu2016,Yamauchi2016,Koyama:2017zmv,Nobukawa:2017ilu}. 
The GDXE comprises three distinct regions: the Galactic center (inner $\sim$100\,pc, or $|l| \lesssim 0.^\circ5$), Galactic bulge (inner $\sim$1\,kpc, or $|l| \lesssim 5^\circ$), and Galactic ridge ($|l| \approx 5-100^\circ$). 
Throughout this paper, we use these spatial definitions to distinguish the Galactic center X-ray emission (GCXE), the Galactic bulge X-ray emission (GBXE), and the Galactic ridge X-ray emission (GRXE).

Recently, the seemingly coherent picture of the GDXE, with the Galactic center, bulge, and ridge all dominated by a similar unresolved binary population, has been challenged by \nustar\ broad-band and \suzaku\ Fe line measurements. 
Previously, both low-energy ($<$10\,keV) and high-energy ($>$10\,keV) observations were attributed to a dominant population of magnetically accreting white dwarf (WD) binaries, in particular intermediate polars (IPs), with an average WD mass of $\langle M_{WD} \rangle \approx 0.5-0.66 M_\odot$~\citep{Revnivtsev2006, Krivonos2007, Yuasa2012, Heard2013A}.
However, \nustar\ observations of the inner 100\,pc now indicate that the Galactic center is dominated by heavier IPs with $\langle M_{WD} \rangle > 0.8 M_\odot$~\citep{Perez2015,Hailey2016,Hong2016}.
In addition, \suzaku\ measurements have revealed differing Fe line intensities in the three regions, indicating a changing underlying source population~\citep{Xu2016,Yamauchi2016,Koyama:2017zmv,Nobukawa:2017ilu}.
The Fe line properties of the bulge, in particular, are well-described by a dominant population of non-magnetic WD binaries~\citep{Nobukawa:2017ilu}, in particular quiescent dwarf novae (DNe).

\begin{figure*}[t]
\begin{centering}
\includegraphics[width=0.9\linewidth, angle=0]{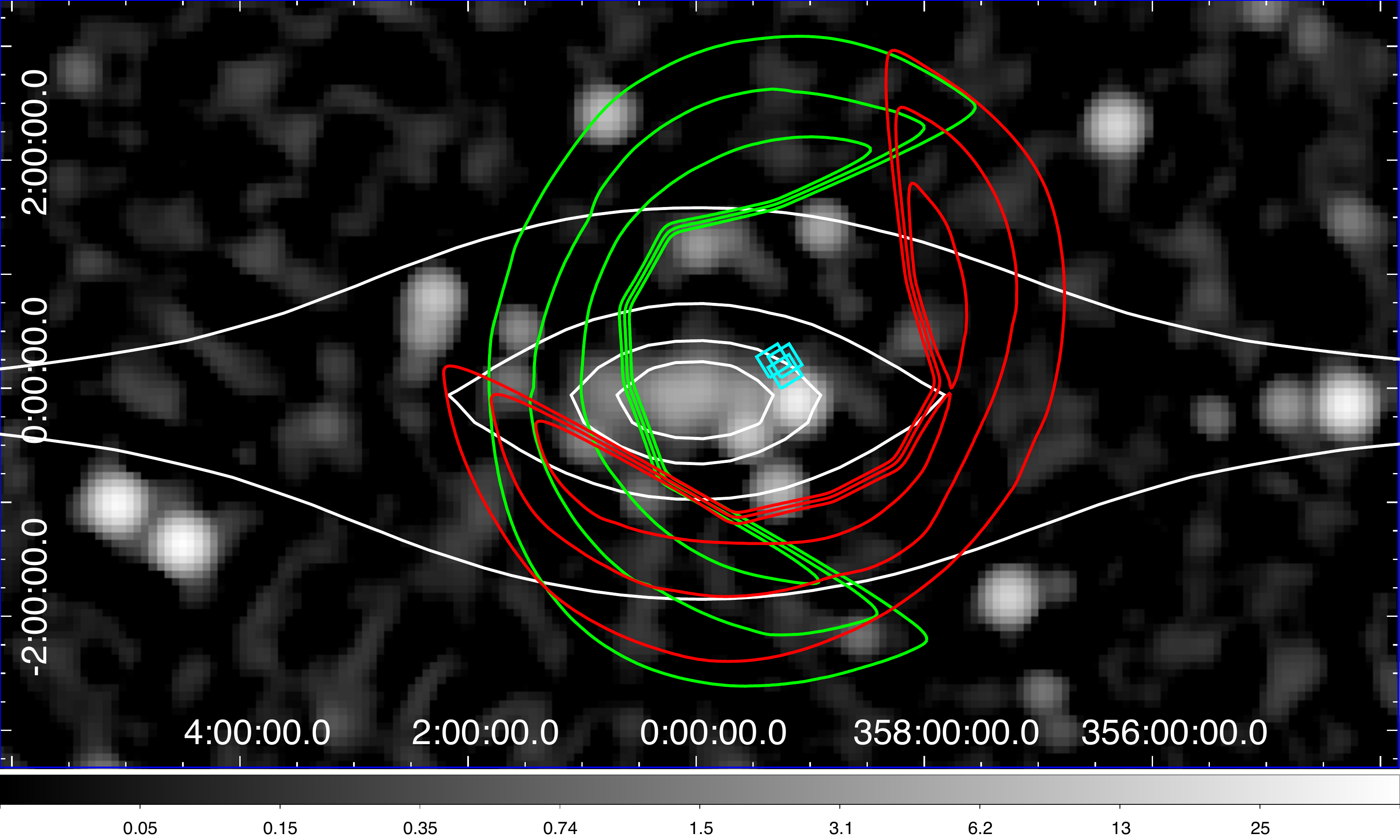}
\vspace{-0.in}
\caption{\label{fig:sky} Sky coverage of 2-bounce photons (cyan squares) and 0-bounce photons from FPMA (red) and FPMB (green) after removal of stray light, bad pixels, and ghost rays, overlaid on the \integral\ (angular resolution 12\arcmin\ FWHM) 17--60\,keV image of the central $12^\circ \times 6^\circ$ of the Galaxy~\citep{Integral14year}. To avoid diffuse emission from the Galactic center, we use only data from FPMA in this analysis. The gray color scale is in units of mCrab. The red and green contours indicate the efficiency due to vignetting effects. The white contours (linearly-spaced) indicate the projected stellar mass density distribution of the Nuclear Stellar Bulge (excluding the Nuclear Stellar Cluster) and the Nuclear Stellar Disk~\citep{Launhardt:2002tx}. }
\end{centering}
\end{figure*}

The \nustar\ telescope~\citep{Harrison2013}, due to its angular resolution of 18\asec full-width half-max ($<1$~pc at the Galactic center) for focused photons and wide aperture (several deg$^2$) for unfocused photons~\citep{Wik:2014boa, Madsen:2017vxr, Perez:2016tcq}, provides the unique ability to separately measure the innermost Galactic center, the bulge, and the ridge \emph{using the same instrument}. 
The goal of this paper is to use the broad-band X-ray energy range of \nustar\ to probe the origin of the diffuse X-ray emission of the inner $\sim1 - 3^\circ$ of the Galactic bulge, and to compare with previous \nustar\ studies of the inner 10\,pc and inner 100\,pc of the Galactic center.
Our focus here is on the underlying point source components of the diffuse emission in both regions;
the \nustar\ Galactic center analyses~\citep{Perez2015,Hailey2016,Hong2016} excluded regions of known truly diffuse emission, such as molecular clouds, and the diffuse emission of the bulge is known to be dominated by unresolved point sources~\citep{Revnivtsev2009,Hong:2012MNRAS}.

This analysis offers the following advantages.
First, 
we are able to search for a possible multi-temperature component of the GBXE.
The low-energy Fe line studies~\citep{Yamauchi2016,Koyama:2017zmv,Nobukawa:2017ilu,Xu2016} that conclude that the bulge is dominated by DNe are in contrast to the IP interpretation favored by previous hard X-ray measurements of the bulge and ridge~\citep{Yuasa2012, Krivonos2007}.
If the bulge is indeed dominated by a soft spectrum consistent with DNe, the broad energy range of \nustar\ allows us to search for evidence of a separate IP population contributing to the hard X-ray band.
Such an observation could indicate that the heavy IP population observed in the central $\sim100$\,pc extends outwards into the bulge. 
Second, 
if a substantial IP population does exist in the bulge, then the \nustar\ high-energy measurement is necessary to derive the correct WD mass, as narrow-band analyses that fit only the cooler regions of the accretion flow can bias the derived shock temperatures (and thus derived WD masses) to lower values~\citep{Hailey2016}.
Finally, 
this work has complementary systematics to the low-energy X-ray studies of the bulge.
In particular, these studies rely on comparison of diffuse Fe line properties with a limited sample of isolated binary systems (3 polars, 16 IPs, 16 DNe, and 4 active binaries), with large spectral variations within each source sample. 
The analysis presented here, in contrast, must exclude the Fe line regions (see Sec.~\ref{sec:binenergy}), and instead bases interpretation on the broad-band continuum shape, which can be better determined than temperatures derived using narrow energy bands. 

In Sec.~\ref{sec:previous}, we review the previous studies of the GDXE and introduce the various candidate source populations, which will motivate the spectral models used in our analysis. 
In Sec.~\ref{sec:dataprep}, we outline the \nustar\ observations used and the data preparation, in particular the use of unfocused (``0-bounce") photons that allows us to have a larger sky coverage than typical for focused (``2-bounce") photons. 
We present the spectral modeling analysis in Sec.~\ref{sec:spec}. 
These results are discussed in Sec.~\ref{sec:disc}, and conclusions are presented in Sec.~\ref{sec:conc}.

\section{Review of possible origins of the GDXE}
\label{sec:previous}

Measurements of the large-scale diffuse morphology and deep measurements of isolated fields indicate that the broad-band nature of the GDXE is primarily due to unresolved low-luminosity X-ray point sources.
\citet{Revnivtsev2006} found that the 3-20\,keV spectrum along the Galactic ridge traces the Galactic stellar mass density (as measured by the infrared surface brightness of the Galaxy), with a measured luminosity consistent with the emissivity per unit stellar mass of X-ray sources in the solar neighborhood. 
We stress that it has a consistent value over the large-scale extent of Galaxy: disk and bulge~\citep{Revnivtsev:2006gf,Krivonos2007}.
Subsequent deep \chandra\ observations of the ``Limiting Window", a low-extinction region in the Galactic bulge, resolved over 80\% of the $\sim6-7$\,keV emisison into discrete sources~\citep{Revnivtsev2009}.
A truly diffuse origin is disfavored, as such a high-temperature ($\sim10^8$\,K) interstellar plasma could not be contained by the gravitational well of the Galactic disk, and no Galactic source capable of supplying the necessary energy replenishment rate has been identified~\citep{Ebisawa2001,Tanaka2002,Ebisawa:2005zt}.

The identity of the dominant point source class, however, remains unclear. 
Cataclysmic Variables (CVs) are a natural candidate, as they have a large space density with low-luminosity objects dominating the overall X-ray flux of the population~\citep{Sazonov2006, Byckling2010, 2012MNRAS.419.1442P, 2013MNRAS.432..570P, Reis:2013bh,Pretorius:2014wma, Britt:2015yca}.
CVs are binary systems in which a WD accretes matter from a late-type main sequence companion via Roche-lobe overflow~\citep{Mukai:2017qww}. 
The most numerous subclass of CVs is the DNe, which in quiescence are thought to produce hard X-rays from the optically thin boundary layer of the accretion disk. 
Polars and IPs, classes of magnetic CVs (mCVs) that have WD magnetic fields strong enough to distort the inner accretion disk, are less numerous, but are especially prolific sources of $\simgt5$\,keV emission.
In contrast, coronally active binaries (ABs), which are normal star binaries with increased X-ray emission due to strong magnetic fields, have been proposed as at most a sub-dominant contribution to the X-ray source population~\citep{Revnivtsev2009}.

Previous studies of the Galactic center, bulge, and ridge, which have favored a uniform IP interpretation of the diffuse emission with an average WD mass of $\langle M_{WD}\rangle \approx 0.49-0.66 M_\odot$, have been challenged by recent \nustar\ observations of the Galactic center and \suzaku\ measurements of the Fe line properties in each region.
Measurements of the ridge by \integral\ showed a hard X-ray (17-60\,keV) luminosity per stellar mass resembling that of mCVs in the solar neighborhood,
with a spectrum consistent with a 1-dimensional accretion flow model of IPs~\citep{Suleimanov2005} with an average WD mass of $\langle M_{WD}\rangle \approx 0.5 M_\odot$~\citep{Krivonos2007}. 
In the bulge, \suzaku\ measured a 2-50\,keV spectrum, which if interpreted as IPs yields $\langle M_{WD} \rangle = 0.66^{+0.09}_{-0.07}M_\odot$~\citep{Yuasa2012}.
The \chandra\ discovery  in the ``Limiting Window" of 10 periodic X-ray sources, interpreted to be mostly mCVs, lent support to this IP interpretation in the bulge~\citep{Hong2012}. 
In the inner 100\,pc, ~\citet{Heard2013A} used \xmm\ observations of the Galactic center to argue that the harder ($>5$\,keV) component of the soft (2-10\,keV) spectrum is dominated by mCVs, primarily IPs, with an average WD mass of $\langle M_{WD} \rangle =0.49 \pm 0.02 M_\odot$. 
In hard X-rays, \citet{Revnivtsev2006} showed that the \rxte\ and \integral\ combined 3-100\,keV spectrum of the Galactic center supports a composite population of IPs, polars, DNe, and ABs, with the $>$10\,keV spectrum dominated by IPs. This result should be viewed with caution, however, as it is now known that the \integral\ $>$20\,keV Galactic center spectrum suffered significant contamination from molecular clouds, non-thermal filaments, and a bright pulsar wind nebula~\citep{Mori:2015vba}. 

The results of the \nustar\ Galactic center survey program, however, have revealed a much heavier average WD mass in the inner $\sim$100\,pc. 
This program discovered an X-ray component in the inner $\sim8$\,pc\,$\times$\,4\,pc that is significantly harder than that previously measured in the Galactic center, bulge, or ridge~\citep{Perez2015}.
Subsequent \nustar\ broad-band measurements of isolated IPs showed that both this central hard X-ray emission and low-energy \chandra\  measurements of the inner $\sim100$\,pc~\citep{Muno2004,Muno2009} indicate a dominant population of IPs with higher average WD mass, $\langle M_{WD}\rangle \simgt 0.8 M_\odot$, than previously derived~\citep{Hailey2016}. 
This mass agrees with the average WD mass in CV systems measured by the wide-field Sloan Digital Sky Survey, $\langle M_{WD}\rangle = 0.83 \pm 0.24 M_\odot$~\citep{Zorotovic2011}. 
Further support comes from the \nustar\ Galactic center point source survey, which showed that the log$N$-log$S$ distributions of both \nustar\ hard X-ray point sources and \chandra\ soft X-ray sources in the same region can be accounted for by a population with average temperature $kT > 20$\,keV, consistent with these heavier WD masses~\citep{Hong2016}.

In contrast to this heavy IP population, an increasing fraction of non-magnetic CVs is indicated with increasing distance from the Galactic center by recent measurements of Fe line properties in the bulge and ridge~\citep{Xu2016,Yamauchi2016,Nobukawa:2017ilu}.
Fe emission lines are sensitive tracers of hot plasma temperature in accretion flow.
\citet{Xu2016} compared \suzaku\ measurements of the Fe line properties of the 2-10\,keV spectrum of the bulge with those of isolated CVs and ABs in the solar neighborhood, showing that the Fe properties of the bulge are inconsistent with the selected sample of IPs or polars, but similar to DNe. 
They conclude that the low-energy emission is dominated by faint DNe and ABs, although mCVs may dominate the high-energy spectrum. 
This result is supported by \citet{Yamauchi2016} and \citet{Nobukawa:2017ilu}, who measured varying Fe line equivalent widths in \suzaku\ spectra of the Galactic center, bulge, and ridge, which were interpreted as changing population fractions in each region. 
Non-magnetic CVs, although originally proposed by \citet{Mukai1993}, had previously been disfavored as a dominant contributor to the GDXE as their typical luminosity was thought to be too high. Recent work has revealed a much lower average luminosity~\citep{Reis:2013bh}, allowing the individual sources to escape the point detection limit~\citep{Ebisawa2001}.

In this paper, we use new \nustar\ measurements of the Galactic bulge and previous \nustar\ measurements of the Galactic center to bolster the case that the diffuse emission in the two regions arises from differing underlying point source populations.
The apparent discrepancy between the average WD mass in different regions of the GDXE could thus be due to misidentification of the dominant source population in the bulge and ridge. 
For the same WD mass, the inner accretion flow of a non-magnetic CV has a lower temperature than the standoff shock and accretion column of a mCV~\citep[e.g.,][]{Ezuka1999, Anzolin2008, Ishida2009, Mukai:2017qww}.
Thus, the softer spectrum of the bulge and the harder spectrum of the center could arise from a population with the same average WD mass, but with different accretion properties.
Alternately, perhaps more massive stellar remnants accumulate in the Galactic center due to dynamical friction~\citep{Portegies2001}, or stellar remnants in binary systems become more massive due to increased accretion rates in the dense environment, although we note that mass gain and mass segregation in different stellar environments is a complex and unresolved topic~\citep{Wijnen2015,Ge2014}.

Additional support for a dominant non-magnetic CV component of the diffuse emission further along the disk comes from the \nustar\ point source survey in the Norma Arm region.
\chandra\ measurements of the log$N$-log$S$ distribution in this region were found to be in very good agreement with that predicted for CV populations~\citep{Fornasini:2014ooa}.
The \nustar\ survey of this region measured a log$N$-log$S$ distribution that is consistent with \chandra\ sources if the average source spectrum is thermal with $kT \approx 15$\,keV~\citep{Fornasini2017}.
This is significantly harder than the $kT \approx 8$\,keV spectrum derived by previous soft X-ray measurements of the Galactic center~\citep{Muno2004,Heard2013A}, but softer than \nustar\ measurements of Galactic center, thus possibly indicating a lower fraction of IPs relative to other types of CVs in this region.

\section{\nustar\ Data Preparation}
\label{sec:dataprep}

\subsection{Introduction to 0-bounce and 2-bounce photons}
\label{sec:0b2b}

To obtain the wide solid-angle aperture necessary to study the diffuse X-ray emission from the Galactic bulge, we exploit \nustar's ability to measure unfocused, or ``0-bounce", X-rays from a $\sim1-3^\circ$-radius field-of-view (FOV). The details of the \nustar\ instrument relevant for this analysis are summarized here.

\emph{NuSTAR} has two identical telescopes, each consisting of an independent optic and focal-plane detector, referred to as FPMA and FPMB. The optics use a conical approximation to the grazing-incidence Wolter-I design, in which X-rays are focused by reflection from an upper parabolic mirror section and then a lower hyperbolic mirror section. 
The FOV for these focused, or ``2-bounce", photons is  $\sim13\arcmin\ \times 13\arcmin\ $. 

To block unfocused X-rays from reaching the \nustar\ detectors, a series of aperture stops are attached to each focal-plane bench. 
However, this shielding is not complete, and there remains a $\sim3.5^\circ$-radius aperture, partially blocked by the optics bench, from which totally unfocused, or ``0-bounce", photons can reach the detectors. 
The solid-angle aperture for 0-bounce photons is over two orders of magnitude larger than the FOV for focused photons, even after accounting for blocking by the optics and vignetting effects due to the aperture stop.

This large solid angle for 0-bounce photons allows us to study the diffuse emission of the Galactic bulge.
In this analysis, we construct a spectrum using photons from the full detector plane, which are a combination of photons from the 0-bounce aperture, the 2-bounce FOV, and instrument background (similar to the technique used in~\citet{Ng:2019gch}).
Via this technique, we lose the imaging capabilities of the \nustar\ optics, but gain spectral information from a large region of the sky. 

Although the aperture for 0-bounce photons is much larger, the 2-bounce FOV is not entirely negligible, due to the increasing intensity of the Galactic diffuse flux in the 2-bounce FOV of our observations. 
Using the mass model of the Nuclear Stellar Disk and Nuclear Bulge, as described in Sec.~\ref{sec:2bounce}, we estimate that the 2-bounce count rate from the Galactic center in our spectrum is $\sim20-25$\% as high as the 0-bounce count rate from the bulge.
We thus need to account for both components in our spectral analysis. 
We correctly account for the different 0-bounce and 2-bounce effective areas, efficiencies, and sky coverage during spectral modeling, as described in Sec.~\ref{sec:obs} and \ref{sec:specprep}. 
(We note that ``1-bounce" photons or ``ghost rays", which undergo only single mirror reflection, have an effective area that is $<$10\% as high as the 2-bounce effective area and are thus negligible~\citep{2017arXiv171102719M}.)

\subsection{Observations, data cleaning, and sky coverage}
\label{sec:obs}

\emph{NuSTAR} observed the Galactic center for a total of $\sim2$~Ms over the period from July 2012 through October 2014~\citep{Mori:2015vba,Hong2016}. 
Using focused photons, these observations imaged the central $\sim1.4^\circ\times0.6^\circ$ of the Galaxy. 
We use three tiled observations (Table~\ref{tab:obs}) from the ``Block B" survey~\citep{Hong2016}, chosen to minimize the 2-bounce flux from bright sources closer to the Galactic center.

We derive fit and flux parameters from each of these three observations individually. We do not stack the three spectra, as each has a different, time-dependent instrumental background. In particular, the normalization of the line and continuum emission components, as well as the spectral shape of the low-energy instrumental background, can vary between each observation. As we are not limited by the statistics of each observations, but instead by systematics associated with the instrumental background model~\citep{Ng:2019gch}, we leave the three observations as independent measurements.

Data reduction and spectral extraction were performed with the \nustar\ Data Analysis Software pipeline (\emph{NuSTARDAS}) v1.5.1. 
We first remove all data from passages through the South Atlantic Anomaly. 
In addition, we remove any periods with a 3--5\,keV count rate higher than $\sim$0.3\,cts/s, as this is indicative of increased background due to solar flaring. 
This removed $\sim25$\,ks from Obs 40032006001.

To study the diffuse emission, we must also remove any contaminating emission from bright isolated point sources. 
This consists of ``stray light", which is extremely bright contamination caused by 0-bounce photons from isolated bright point sources within $\sim3^\circ$~\citep{2017arXiv171102719M}.  
We flag as ``bad" any pixels with this contamination. These are then removed during the data screening procedure implemented in \emph{NuSTARDAS}. 
We discuss in Sec.~\ref{sec:lumi} the upper luminosity limit on a 0-bounce source that is just below the threshold for stray light removal, and which therefore may remain after this cleaning procedure.
Using the \nustar\ Galactic center source catalog~\cite{Hong2016}, the total 10-40\,keV flux of identified point sources in our 2-bounce FOV is $\sim10^{-6}$~ph~s$^{-1}$~cm$^{-2}$. This is negligibly small compared to our measured GBXE flux.

To construct a sky-exposure map for 0-bounce photons, we use the instrument geometric model from the {\tt nuskybgd} code~\citep{Wik:2014boa}. 
This map is corrected for obscuration by the optics bench and the vignetting effect due to the aperture stops attached to each focal plane bench. 
The effective solid angle of sky coverage for each observation, after accounting for these effects, is given in Table~\ref{tab:obs}.
Note that this pixel removal affects the average solid angle, as different detector areas see slightly different regions of sky with a different efficiency.

The resulting total sky coverage of these three observations is shown in Fig.~\ref{fig:sky}. 
Although FPMA and FPMB have largely overlapping FOVs for focused photons, their 0-bounce apertures cover very different regions of the sky, as indicated in the green and red contours of Fig.~\ref{fig:sky}.
We use only data from FPMA for this analysis, as the 0-bounce sky coverage of FPMA covers the Galactic bulge while avoiding the inner $\sim$100\,pc Galactic center region.

\begin{table*}
\caption{\label{tab:obs} \emph{NuSTAR} observations used for this analysis. Note that only data from FPMA is used, in order to minimize contribution from the inner $\sim$100\,pc Galactic center region.}
\begin{ruledtabular}
\begin{tabular}{ccccccc}
\multicolumn{1}{c}{Observation ID} & \multicolumn{2}{c}{Pointing (J2000)\footnote{Roll angle was 332$^\circ$ for all.} } & Effective Exposure\footnote{After all data cleaning.} & Effective Detector Area\footnote{After stray light, ghost ray, and bad pixel removal.} & Effective Avg. Solid Angle\footnote{Average solid angle of sky from which 0-bounce photons can be detected, after correcting for removal of stray light and bad pixels, as well as efficiency due to vignetting effects.} \\
\multicolumn{1}{r}{} 		& RA (deg) & DEC (deg) & FPMA (ks)  & FPMA (cm$^2$)       & FPMA (deg$^2$)  \\
\hline
\multicolumn{1}{r}{40032002001}  				& 265.7969 & $-29.5139$ & 39.8 & 7.14 & 4.06 \\
\multicolumn{1}{r}{40032003001} 					& 265.6991 & $-29.4613$ & 39.8 & 8.18 & 3.47 \\
\multicolumn{1}{r}{40032006001} 					& 265.7595 & $-29.3762$ & 28.6 & 9.98 & 3.76 \\
\end{tabular}
\end{ruledtabular}
\end{table*}

\begin{figure*}[t]
\vspace{-0.1in}
   \begin{centering}
   \subfloat{%
        \includegraphics[width = 0.68\linewidth]{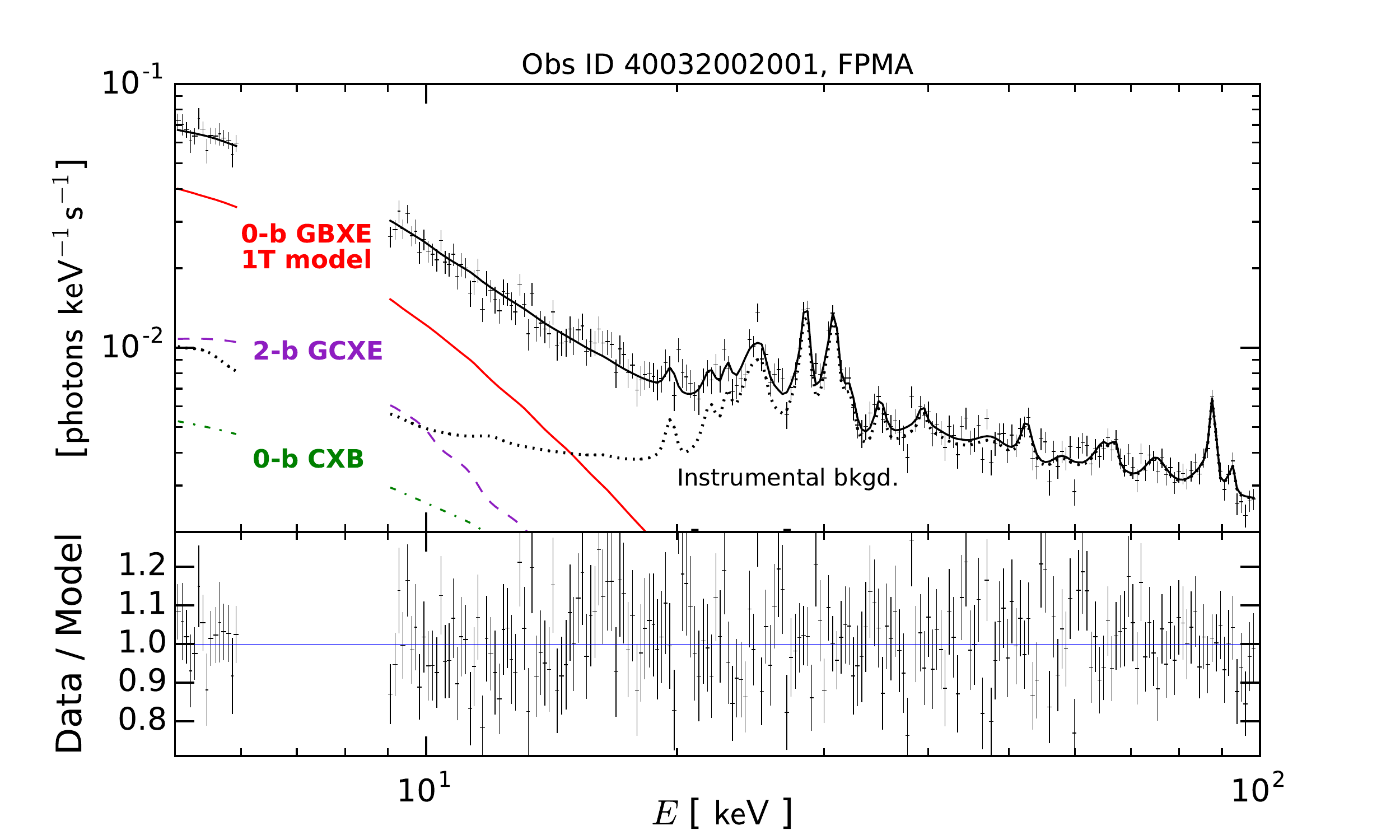}               
    }\qquad
   \subfloat{%
        \includegraphics[width = 0.68\linewidth]{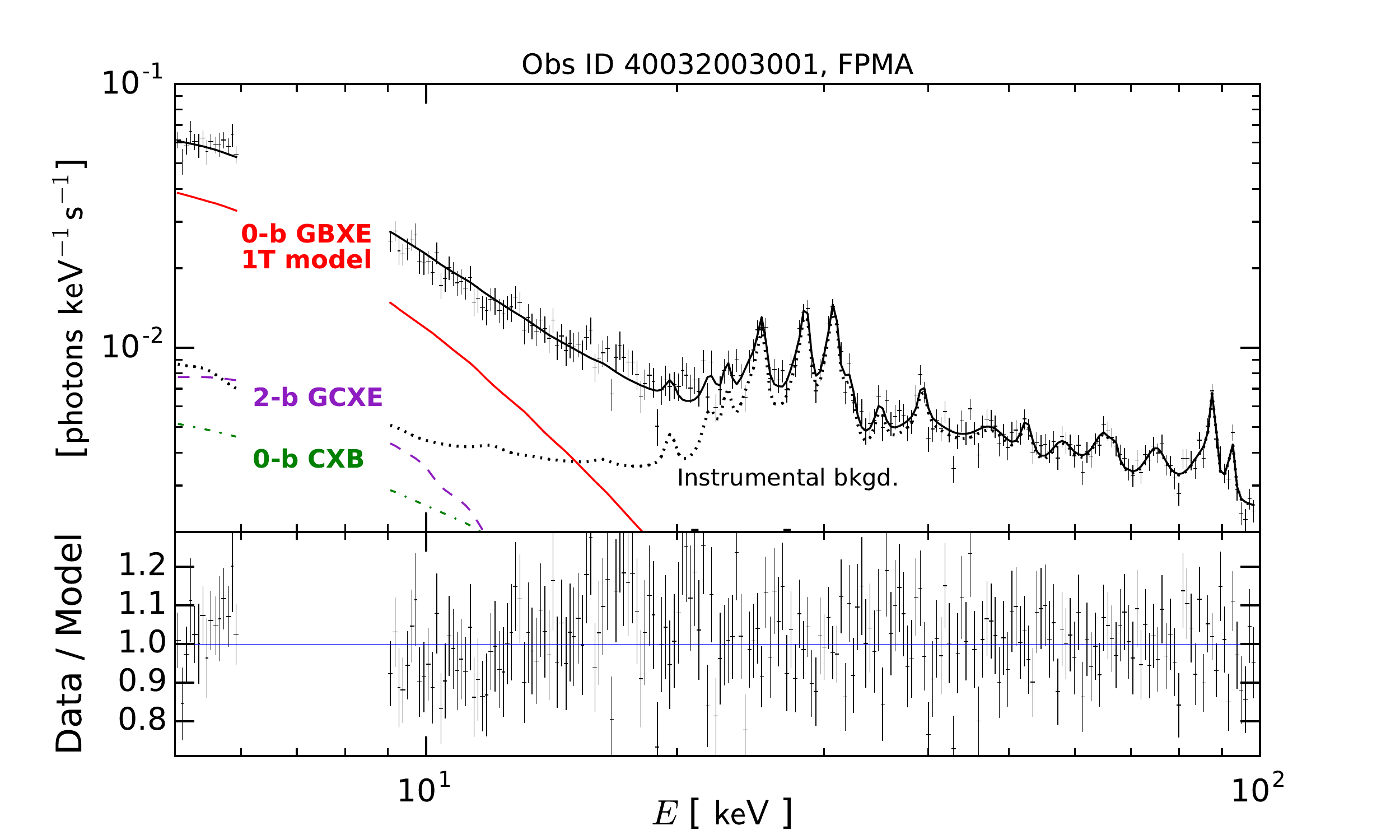}               
    }\qquad
    \subfloat{%
        \includegraphics[width = 0.68\linewidth]{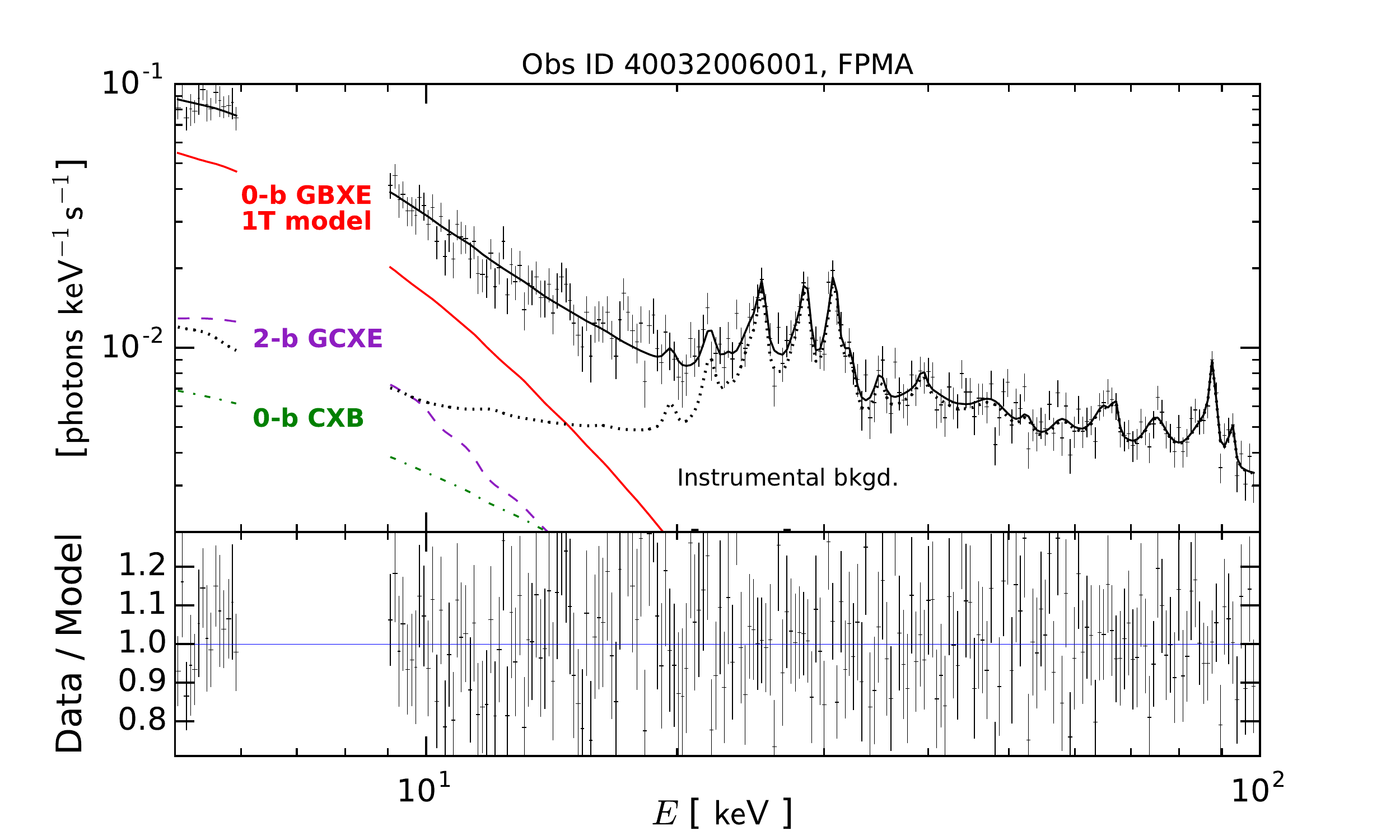}               
    }\qquad
    \vspace{-0.1in}
   \caption{\label{fig:1T} Data and folded best-fit model spectra with the 0-bounce GBXE described by the 1T model, for FPMA of 40032002001 (top),  40032003001 (middle), and 40032006001 (bottom). Model components include the 0-bounce GBXE~(red, solid) described as a 1-temperature thermal plasma, and the 2-bounce GCXE~(purple, dashed), 0-bounce CXB (green, dash-dot), 2-bounce CXB (below y-axis range) and detector background~(gray, dotted) as described in the text. The ratio of the data to the best-fit model is shown in the bottom panel. All errors shown are 1$\sigma$ statistical errors. 
}
\end{centering}
\end{figure*}

\subsection{Spectral file preparation}
\label{sec:specprep}

We construct a spectrum using photons from all  portions of the detector not contaminated by stray-light.
Our spectral model must account for five components: 0-bounce diffuse emission from the Galactic bulge, 2-bounce diffuse emission from the Galactic center, both 2-bounce and 0-bounce emission from the cosmic X-ray background (CXB), and instrumental background. 

To account for both 0-bounce and 2-bounce astrophysical emission within the same spectra, we make custom auxiliary response files (ARFs) for each observation and each model component that account for the differing effective area, in units of cm$^2$, and effective average solid angle of sky coverage, in units of deg$^2$. 

For the 2-bounce spectral components, we first derive the \nustar\ effective area file for the detector region after all bad pixel and data cleaning using the standard {\tt numkarf} procedure from the \emph{NuSTARDAS}.
This accounts for the effective area of the focusing optics, as well as efficiencies due to detector absorption.
We scale this file by the corresponding 2-bounce sky coverage, to yield an ARF file in units of cm$^2$\,deg$^2$. 

For the 0-bounce spectral components, 
the effective area is the physical detector area, after removal of stray light and bad, pixels, listed in Table~\ref{tab:obs}. 
We scale this by the effective sky coverage of 0-bounce photons, after all data cleaning and accounting for the efficiency due to vignetting effects, also listed in Table~\ref{tab:obs}. 
These model components are also subject to the energy-dependent efficiency for photons to pass through the beryllium window in front of the detectors, which drops sharply below 10\,keV but is approximately unity elsewhere, and an absorption term ({\tt nuabs},~\citet{Madsen:2017vxr}) that accounts for the detector CdZnTe dead layer. (This beryllium widow efficiency and detector absorption are included for the 2-bounce components via the effective area file produced by {\tt numkarf}.)

The derived (or fixed) fluxes of all astrophysical components are thus reported in units of $\rm ph\,s^{-1}\,cm^{-2}\,deg^{-2}\,keV^{-1}$. 
Spectral fitting and flux derivations are performed in XSPEC version 12.9.0~\cite{Arnaud1996}.

\subsection{Binning and Energy Range}
\label{sec:binenergy}

For spectral analysis, we use the energy range 5--100\,keV. 
The highest-energy ($>50$\,keV) region is necessary to properly constrain the continuum component of the instrumental background. 
In contrast, the low-energy (3--5\,keV) components of the instrumental background cannot be well constrained by our data, and are thus ignored. 
In particular, this instrumental background includes line-like emissions at $\sim3.5$\,keV and $\sim4.5$\,keV with strengths that vary between different observations~\citep{Wik:2014boa}, and which can significantly bias the derived low-energy continuum shape (for detailed discussion of the difficulty of using the 3--5\,keV region with the 0-bounce analysis technique, see~\citet{Ng:2019gch}). 

Additionally, we ignore the energy range 6--9\,keV, as this region contains significant line emission from neutral and ionized Fe that is degenerate between the 2-bounce Galactic center diffuse component and the 0-bounce Galactic bulge diffuse component~\citep{Nobukawa:2017ilu}.   
These Fe line properties are a valuable input to constraining the temperature of any emission. However, any observed line flux is a combination of 0-bounce emission from the bulge and 2-bounce emission from the center. As the IPM model that we use to describe the Galactic center emission (see Sec.~\ref{sec:2bounce}) does not describe this line emission, we cannot accurately model how much of the observed flux is due to the bulge emission only. 

Instead, the focus of this paper is to use the broad-band continuum to constrain the spectrum of the bulge emission. For this reason, we still include 5--6\,keV, as this gives valuable constraints on the continuum shape. 
This study complements the recent Fe line studies, as it allows us to search for a high-energy contribution to the population.
 
We bin each spectrum using a logarithmic binning scheme, with 200 bins per decade in energy.
This allows for clear visual inspection when viewing a large energy range on a logarithmic axis, 
and yields a statistical error that is approximately uniform at $\sim$10\% across the full energy range, with $\sim100$ counts per bin at the lowest energies and at least 50 counts per bin at the highest energies. 
There are no significant variations in the derived Galactic diffuse fit parameters due to alternate binning schemes. 

\begin{table*}[ht]
\caption{\label{tab:modelfits} Summary of best-fit parameters for the 1T and IPM models of the Galactic bulge X-ray continuum. All errors are 90\% C.I. Flux is quoted over the energy range 5--20\,keV. The column density is fixed to $N_H = 3.0\times10^{22}$\,cm$^{-2}$ for all. Also listed are the power-law index and normalization relative to the internal continuum of the instrumental background model described in the text.}
\begin{ruledtabular}
\begin{tabular}{lllllll}
\multicolumn{1}{c}{} & \multicolumn{2}{c}{40032002001} & \multicolumn{2}{c}{40032003001} & \multicolumn{2}{c}{40032006001}\\
\multicolumn{1}{l}{Parameter} & \multicolumn{1}{c}{1T} & \multicolumn{1}{c}{IPM} & \multicolumn{1}{c}{1T} & \multicolumn{1}{c}{IPM} & \multicolumn{1}{c}{1T} & \multicolumn{1}{c}{IPM}\\
\hline
$\Gamma_{internal}$ & \multicolumn{2}{c}{ 1.7$^{+0.2}_{-0.3}$ } & \multicolumn{2}{c}{1.6$^{+0.3}_{-0.3}$ } & \multicolumn{2}{c}{1.5$^{+0.3}_{-0.3}$ } \\
Rel. Norm. & \multicolumn{2}{c}{ 49$^{+20}_{-19}$ } & \multicolumn{2}{c}{36$^{+26}_{-16}$ } & \multicolumn{2}{c}{35$^{+27}_{-16}$ } \\
\hline
$kT$ [keV]  & 8.0$^{+0.3}_{-0.6}$ &  -  & 8.0$^{+0.8}_{-0.7}$ &  -  & 7.4$^{+0.9}_{-0.9}$ &  - \\
$M_{WD}$ [$M_\odot$] & - &  0.44$^{+0.04}_{-0.03}$   &  -  &  0.45$^{+0.04}_{-0.04}$  & - &  0.41$^{+0.05}_{-0.01}$ \\
$F_X$ $[10^{-11}$\,erg\,s$^{-1}$\,cm$^{-2}$\,deg$^{-2}]$ & 10.3$^{+0.6}_{-0.1}$ & 10.0$^{+0.7}_{-0.1}$ & 10.2$^{+0.6}_{-0.2}$ &  10.3$^{+0.7}_{-0.1}$  & 10.4$^{+0.5}_{-0.5}$ &  10.5$^{+0.4}_{-0.6}$ \\
$\chi^2_\nu$ (dof)  & 1.11 (197) & 1.08 (198)   & 1.10 (197) &  1.06 (198) & 1.05 (197) &  1.04 (198)  \\
\end{tabular}
\end{ruledtabular}
\end{table*}

\section{Spectral Model}
\label{sec:spec}

\subsection{Instrumental Background Model}
\label{sec:bkgd} 

A standard \nustar\ instrumental background model has previously been derived from phenomenological fits to ``blank sky" observations~\citep{Wik:2014boa}. 
It is dominated at high energies by a series of Lorentzian lines and a relatively flat continuum component, modeled as a broken power-law with a break at 124~keV. 
At low energies, it is dominated by a 3.5\,keV line,  a 4.5\,keV line, and a $\sim$1\,keV thermal plasma component (the {\verb=apec=} model in {\verb=XSPEC=}) that is possibly due to reflected solar X-rays, whose intensity can vary with time. 

For this analysis, however, we use a slightly modified version of the standard low-energy instrumental background model. 
This is motivated by measurements of occulted data, which is collected during the same observation periods we use for our main analysis but when the FOV of \nustar\ is blocked by the Earth. 
The occulted data gives a better constraint on the internal background spectrum than, for example, blank-sky data collected in another time period, because it is free of low-energy emission from the CXB, which can obscure the low-energy instrumental component, and because it accurately describes instrumental background components that may have long-term time variation.
We find that for energies above $\sim$20\,keV, the occulted data is well described by the above model.
Below $\sim$20\,keV, however, the occulted data indicates residual emission that is not accounted for by this default model. 

These low-energy occulted fits are improved if we replace the $\sim$1\,keV thermal plasma component with a power-law continuum. 
In this modified instrumental background model, we use occulted data from each observation to derive a best-fit power-law index and relative normalization with respect to the high-energy continuum. 
We then freeze both of these parameters in the instrumental background model that we apply to non-occulted data. 
These parameters are shown in Table~\ref{tab:modelfits}.
This modified background model has been validated on extragalactic observations, where it yields the correct expected CXB spectral shape and flux (see Appendix~\ref{app:cxb}).

The instrumental background model we use in this analysis thus consists of 
\emph{(i.)} the continuum and line components from ~\citet{Wik:2014boa}, with the line widths and energies fixed and the continuum and line normalizations left free, and
\emph{(ii.)} the low-energy power-law component described above. 
Compared to the default instrumental model, this low-energy power-law background model has no significant effect on the derived continuum shape of the Galactic bulge diffuse emission, with the derived temperature or IP mass of the GBXE remaining unchanged, within the 90\% C.I., between the two instrumental models.
However, it does affect the measured 5--20\,keV flux of this diffuse emission, lowering it by up to $\sim$15\% as compared to the default model. 
We further discuss uncertainties on the derived GBXE flux in Sec.~\ref{sec:lumi}.

\subsection{Cosmic X-ray Background (CXB) Model}
\label{sec:CXB} 

We model both the 0-bounce and the 2-bounce components of the CXB as a cutoff power-law, with 3--20\,keV flux and spectral indices fixed to those measured by \integral~\citep{Churazov:2006bk}. 
We have verified these spectral parameters and flux with \nustar\ extragalactic data, accounting for the 0-bounce and 2-bounce emission components as described in Sec.~\ref{sec:specprep} and using the instrumental background model described in Sec.~\ref{sec:bkgd}.
This provides further validation of our combined 0-bounce and 2-bounce spectral analysis procedure, and of our modified instrumental background model. 
See Appendix~\ref{app:cxb} for details.

\subsection{2-bounce Diffuse Emission: The Galactic Center}
\label{sec:2bounce} 

It is not possible to independently fit the 2-bounce diffuse emission from the Galactic center and the 0-bounce diffuse emission from the Galactic bulge, due to their similar spectral shapes.
Instead we use previous measurements to fix the 2-bounce spectral shape and relative flux. 

Motivated by \nustar\ measurements of diffuse emission in the inner $\sim$10\,pc~\citep{Perez2015,Hailey2016} and of point sources in the inner $\sim$100\,pc~\citep{Hong2016}, we fix the spectrum of the 2-bounce emission from the Galactic center to a 1-dimensional accretion flow model of IPs (see Sec.~\ref{sec:ipmass}) with $M_{WD} = 0.9 M_\odot$. 
We do not include the soft $kT \approx 1$\,keV component of GCDE, as it is negligible in the energy range  $>5$\,keV~\citep{Perez2015,Hailey2016}.
We note that there is no significant variation in the derived Galactic bulge luminosity or spectral shape (with temperature, IP mass and 5--20\,keV flux remaining consistent within 90\% C.I.) if we instead use a model with $M_{WD} = 0.5 M_\odot$, as has been motivated by the previous studies described in Sec.~\ref{sec:previous}. 
This is due to the overall continuum shape being dominated by the GBXE, with the GCXE only a sub-dominant component. We further discuss the small effects of varying the 2-bounce Galactic center spectral shape on the derived Galactic bulge continuum shape in Sec.~\ref{sec:1Tmodel} and~\ref{sec:ipmass}. 

We use a mass model of the Nuclear Stellar Disk and the Nuclear Bulge~\citep{Launhardt:2002tx} to fix the flux of this component relative to that of the Galactic bulge diffuse component. 
\rxte\ measurements have shown that the 3--20\,keV diffuse X-ray flux traces the stellar mass distribution~\citep{Revnivtsev2009}. 
We thus fix the 3--20\,keV 2-bounce flux relative to the 3--20\,keV 0-bounce flux, in units of erg\,s$^{-1}$\,cm$^{-2}$\,deg$^{-2}$ to be the ratio of the integrated stellar mass in the 2-bounce FOV to that in the 0-bounce FOV.
This ratio is 2.7, 2.2, and 2.5 for Obs 40032002001, 40032005001, 40032006001, respectively. Note that although the flux in units of erg\,s$^{-1}$\,cm$^{-2}$\,deg$^{-2}$ is expected to be higher for the 2-bounce Galactic center component than for the 0-bounce bulge component, the overall count rate of our measurement, in units of ph\,s$^{-1}$, is much higher for the 0-bounce component due to the much larger 0-bounce FOV.
Also, although the ratios differ between the three observations, the 2-bounce sky regions are overlapping and the 0-bounce sky regions are essentially the same; the difference is due to the different stray light and bad pixel removal, and thus different 2-bounce and 0-bounce effective areas and average solid angles, between the three observations.

This component is subject to absorption by the interstellar medium, with interstellar abundances as defined in~\cite{2000ApJ...542..914W},  photoionization cross-sections as defined in~\cite{1992ApJ...400..699B,0004-637X-496-2-1044}, and column density fixed to $6.5\times10^{22}$\,cm$^{-2}$~\citep{Launhardt:2002tx}.

\subsection{0-bounce Diffuse Emission: The Galactic Bulge}

We investigate two different models to describe the diffuse emission from the Galactic bulge:
a single-temperature thermal plasma (1T) model, which allows us to describe the spectrum without assuming the identity of the underlying population, and an intermediate polar mass (IPM) model, which allows us to directly compare with previous \nustar measurements from the Galactic center and \suzaku, \rxte, and \integral measurements of the bulge and ridge. 

\subsubsection{Single-temperature Thermal Plasma Model}
\label{sec:1Tmodel} 

For the same WD mass, non-magnetic CVs exhibit significantly softer emission and cooler average plasma temperatures than mCVs, due to the different mechanisms by which the plasma is heated in each system~\citep{Xu2016,Hailey2016}. 
In non-magnetic CVs, the observed X-ray emission is dominated by hot plasma at the inner boundary of the accretion disk, with approximately half of the gravitational energy of the accreting material dissipated in a relatively cool disk that does not affect the X-ray flux.
In mCV systems, the accretion disk is disturbed by the WD's magnetic field lines, funneling material along these field lines toward the WD poles. 
The hottest material, and thus the hardest X-rays, originate at the stand-off shock region, with a temperature that scales with the WD mass. 
Below this, a column of cooling material with peak emissivity in soft X-rays extends towards the WD surface. 
 
The simplest model for both categories of systems is a one-temperature plasma.
Though this model only approximates the possible multi-temperature nature of the accretion flow, 
it allows us to compare with previous GDXE measurements. 
As with the 2-bounce GCDE, we do not include the soft $kT \approx 1$\,keV component of 0-bounce GBDE, as it is negligible in the energy range  $>5$\,keV~\citep{Yuasa2012}.
This single-temperature plasma is described using the {\tt bremss} model in XSPEC. 
This component is subject to absorption by the interstellar medium, with a column density fixed to the approximate average value in the region covered, $3.0\times10^{22}$\,cm$^{-2}$~\citep{Launhardt:2002tx}.
As we only use data in the range $>5$\,keV, our derived spectral parameters are not significantly affected by changes in this value.

The results of fitting this model to our three observations are shown in Fig.~\ref{fig:1T} and Table~\ref{tab:modelfits}. 
All three spectra are well described by this model, with $\chi^2/d.o.f. \approx 1.1$ and a best-fit temperature of $kT \approx 8$\,keV.

This is significantly softer than the spectrum measured by \nustar\ in the Galactic center, where the point source and diffuse populations are consistent with an average temperature of $kT > 20$\,keV~\citep{Hong2016, Perez2015}, as we discuss further in Sec.~\ref{sec:disc}. 
This conclusion is robust to variations in the assumed 2-bounce Galactic center diffuse spectrum and the low-energy internal background model. 
If we replace the GCDE model with a softer population of IPs with $\langle M_{WD}\rangle \approx 0.5 M_\odot$, as motivated by studies of the GCDE prior to \nustar\ (see Sec~\ref{sec:previous}), the best-fit temperatures increase slightly, by $\sim$10--13\% for each observation, but remains consistent within the 90\% C.I. errors quoted in Table~\ref{tab:modelfits}. Replacing the low-energy internal power-law component with the  $\sim$1\,keV thermal plasma component from the standard \nustar\ background model has a similar effect.

\subsubsection{IP Mass Model}
\label{sec:ipmass}

In order to allow direct comparison of our continuum spectral shape with previous \nustar\ measurements of the GCXE and with previous \suzaku, \rxte, and \integral\ broad-band measurements of the bulge and ridge that favored an IP interpretation~\citep{Revnivtsev2006, Krivonos2007, Yuasa2012}, we also model our GBXE spectrum using an IP mass model.
As most IPs have higher mass transfer rates than polars, and thus significantly higher hard X-ray luminosities, they are a natural candidate for the origin of the GDXE. 
As mentioned above, the X-ray emission of mCVs is an inherently multi-temperature process, and measurements using only the low-energy X-ray spectra will be biased to a lower temperature and mass.
However, broad-band X-ray measurements, such as possible with \nustar\, have been shown to accurately constrain the WD mass~\citep{Hailey2016,Shaw:2018uxl}. 

To model IPs, we use a 1-dimensional accretion flow model that accounts for the density and temperature profile of the accretion column, but uses simple emissivity profile that ignores soft X-ray emission lines~\citep{Suleimanov}. 
The hottest temperature scales with the depth of the potential well, and thus the WD mass. 
We use this IPM model to derive the average WD mass implied by our Galactic bulge spectrum. 
Of course, this mass is only reliable if the spectrum is dominated by emission from IPs, and would be different if attributed to a different point source population, such as non-magnetic CVs. 
As in the 1-temperature model, we include absorption by the interstellar medium with a column density fixed to $3.0\times10^{22}$\,cm$^{-2}$~\citep{Launhardt:2002tx}.

\begin{figure*}
   \begin{centering}
   \subfloat{%
        \includegraphics[width = 0.68\linewidth]{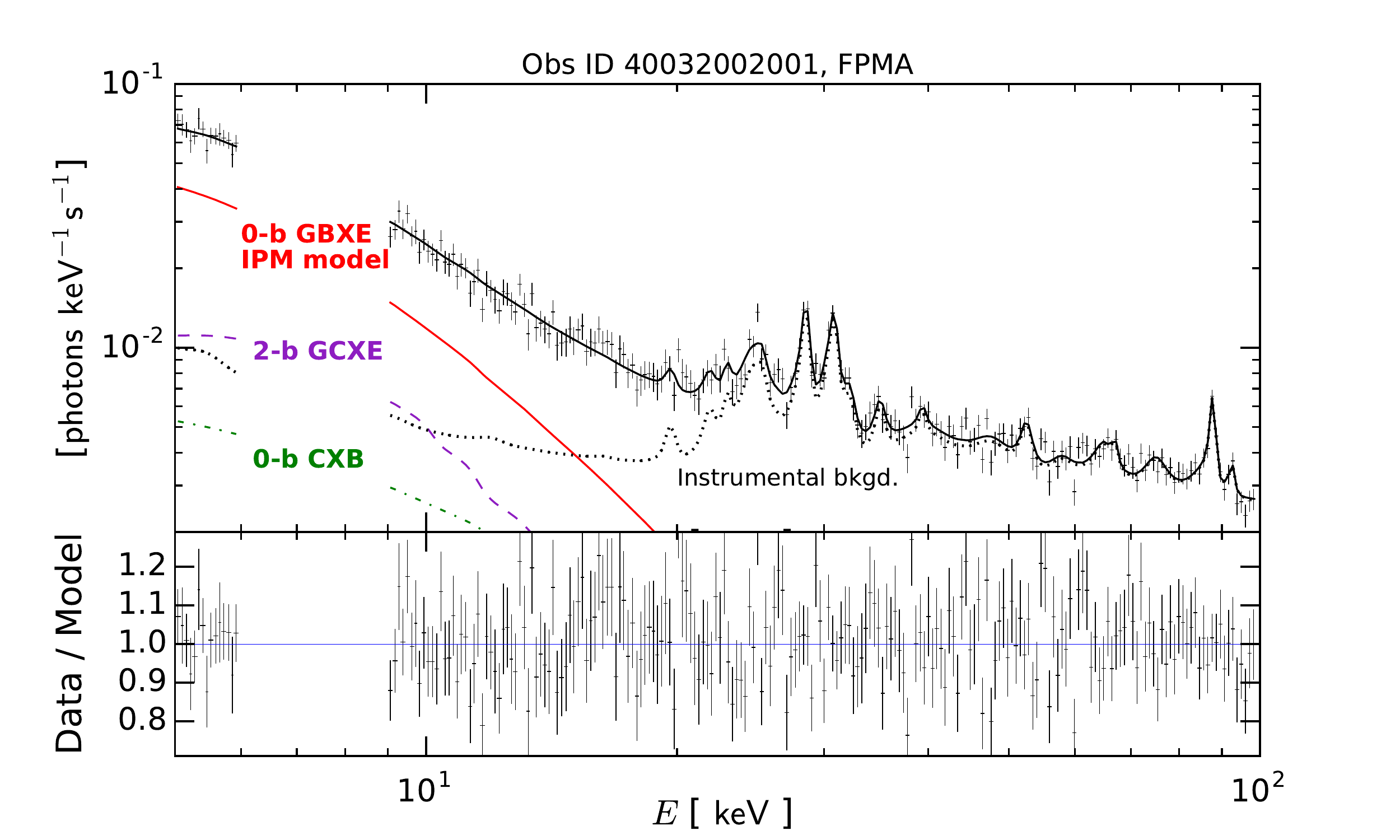}               
    }\qquad
   \subfloat{%
        \includegraphics[width = 0.68\linewidth]{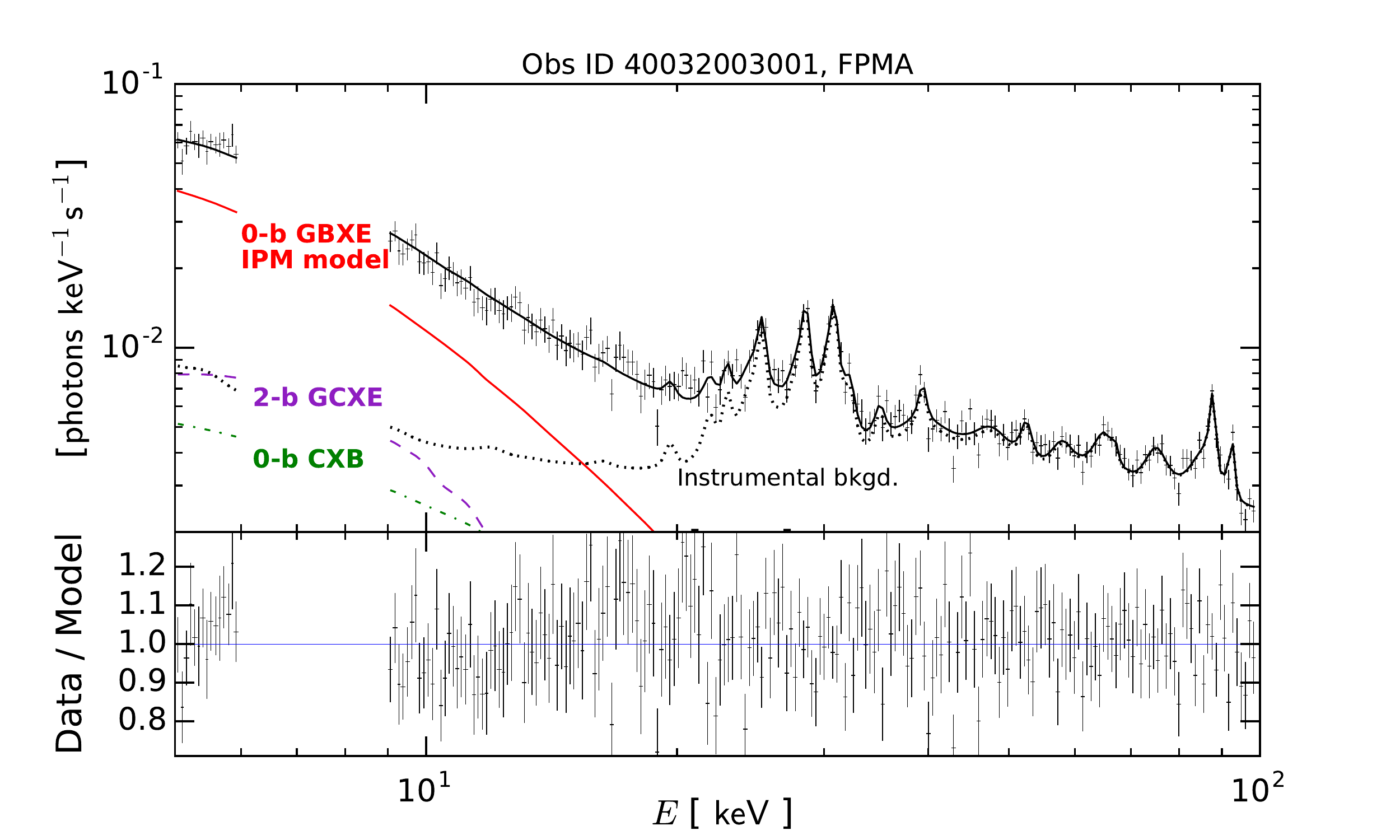}                 
    }\qquad
   \subfloat{%
        \includegraphics[width = 0.68\linewidth]{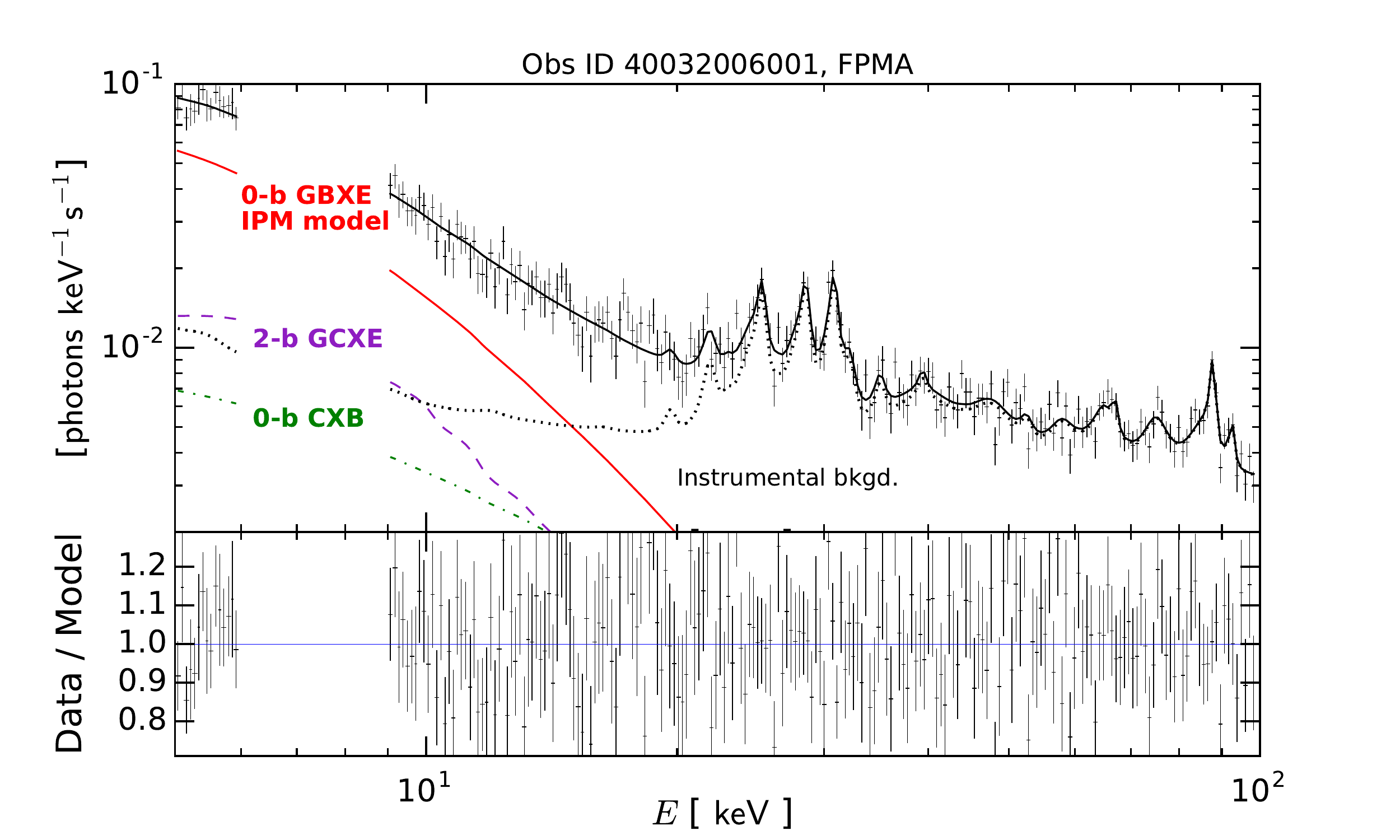}               
    }\qquad
    \caption{\label{fig:IPM} Same as Fig.~\ref{fig:1T}, but with the 0-bounce GBXE described by the IPM model.  
}
     \end{centering}
\end{figure*}

\begin{figure}
\includegraphics[width = 1.0\columnwidth]{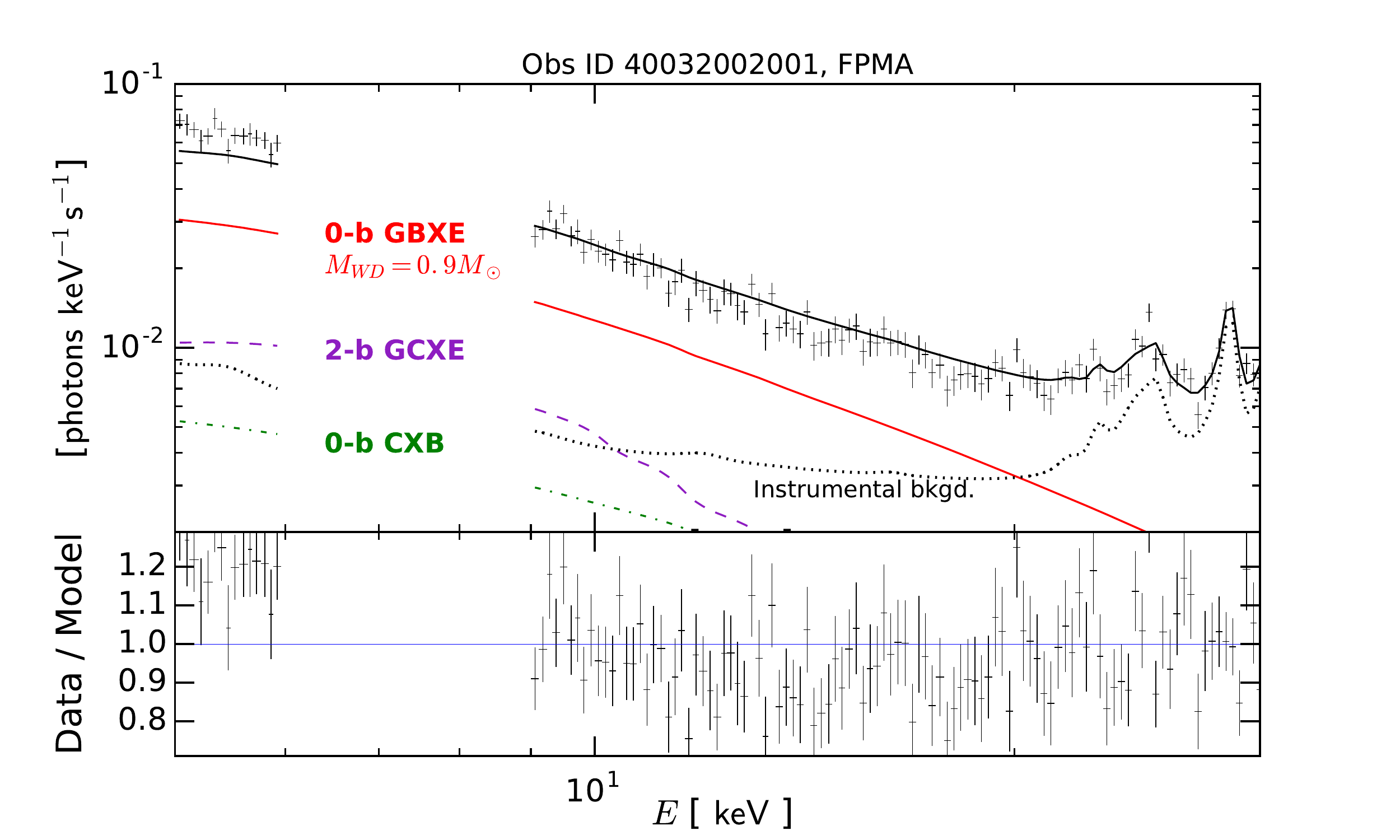}
\vspace{-0.1in}
\caption{\label{fig:IPMzoom} Data and folded best-fit model spectra for FPMA of 40032002001 with the 0-bounce GBXE described by an IPM model with $M_{WD} = 0.9 M_\odot$, zoomed into the energy range 5--25\,keV. Model components as labelled in Fig.~\ref{fig:1T}. Such a heavy average WD mass causes a poor fit to the $E<20$\,keV data.
}
\end{figure}

The results of fitting the IPM model to our three observations are shown in Fig.~\ref{fig:IPM} and Table~\ref{tab:modelfits}. 
Similar to the 1T model, all three spectra are well-described with $\chi^2/d.o.f. \approx 1.1$. 
The best-fit average WD mass of $\langle M_{WD}\rangle \approx 0.4-0.5 M_\odot$ again indicates that the spectrum is significantly softer than that measured by \nustar\ in the Galactic center. 
As with the 1T model, the derived average WD mass increases slightly if we use the softer $\langle M_{WD}\rangle \approx 0.5 M_\odot$ model of the GCXE or replace the low-energy internal power-law component with the usual $\sim$1\,keV thermal plasma component, but remains consistent with the 90\% C.I. errors of Table~\ref{tab:modelfits}. 
In addition, if we instead freeze the WD mass in our IPM model of the 0-bounce bulge emission to be $\langle M_{WD}\rangle = 0.9 M_\odot$, the fit deteriorates significantly for energies below 20\,keV, with a $\chi^2/d.o.f. \approx 1.6$ for each observation. 
This is illustrated in Fig.~\ref{fig:IPMzoom}.

\subsubsection{GBXE Luminosity}
\label{sec:lumi}

The measured 5--20\,keV flux of the GBXE component is presented in Table~\ref{tab:modelfits}. 
In this section, we discuss our flux and luminosity results in the energy band 3--20\,keV, in order to better compare with previous hard X-ray studies of the bulge and ridge.
We note, however, that this is not an exact comparison, as some sub-dominant fraction of the total 3--5\,keV diffuse emission is attributed to the soft $kT \approx 1$\,keV diffuse component~\citep{Perez2015,Hailey2016,Yuasa2012}, which we ignore in our fits.

The measured GBXE flux, $F_{3-20\,keV} \approx 1.4 \times 10^{-10}$\,erg\,s$^{-1}$\,cm$^{-2}$\,deg$^{-2}$, is consistent for each of the three observations.
Integrating over the mass model of~\citet{Launhardt:2002tx} and accounting for efficiency effects due to vignetting, we obtain a luminosity to stellar mass ratio of $L_{3-20\,keV}/M \approx 3.4 \times 10^{27}$\,erg\,s$^{-1}$\,$M_\odot^{-1}$. The dominant uncertainty on this measurement is due to the overall \nustar\ calibration uncertainty, which is $\sim$10\%~\citep{Madsen:2015jea}.
Here we assume a uniform distance to the bulge of 8.5\,kpc~\citep{Revnivtsev2006}.

We estimate the upper luminosity limit at which a 0-bounce point source would remain unresolved in our measurement using two methods. 
First, we estimate the source luminosity that would produce a count rate across our detector that is approximately twice the instrumental background rate in the energy range 3--20\,keV. 
This is the source brightness that should produce a clearly visible stray light pattern on our detectors. 
This yields an estimated upper limit of $F_{3-20\,keV} \approx 5 \times 10^{-11}$\,erg\,s$^{-1}$\,cm$^{-2}$ or $L_{3-20\,keV} \approx 4 \times 10^{35}$\,erg\,s$^{-1}$, which is $<$10\% of our total measured bulge luminosity.
Second, we check this estimate using the \integral\ point source catalog of this region~\citep{Integral9year,Integral14year}. There are three sources that remain after all our data cleaning, the brightest of which has $F_{3-20\,keV} \approx 5.2 \times 10^{-11}$\,erg\,s$^{-1}$\,cm$^{-2}$, consistent with the upper limit estimated via our first method. 
We note that our measured flux is not contaminated by the total flux of this source, due to vignetting-induced inefficiency effects. 

We can compare our luminosity to previous bulge measurements.
Our measured emissivity, $L_{3-20\,keV}/M \approx (3.4 \pm 0.3) \times 10^{27}$\,erg\,s$^{-1}$\,$M_\odot^{-1}$ is consistent with that measured by \rxte\ in the bulge and ridge, $L_{3-20\,keV}/M \approx (3.5 \pm 0.5) \times 10^{27}$\,erg\,s$^{-1}$\,$M_\odot^{-1}$~\citep{Revnivtsev2006},
as well as more recent \rxte\ measurements which give $L_{2-10\,keV}/M \approx (3.0 \pm 0.3) \times 10^{27}$\,erg\,s$^{-1}$\,$M_\odot^{-1}$~\citep{RevMolkov2012}. 
Our measured $L/M$ is also consistent with the 3--20\,keV X-ray emissivity of sources in the stellar neighborhood, $(5.3 \pm 1.5) \times 10^{27}$\,erg\,s$^{-1}$\,$M_\odot^{-1}$, excluding young coronal stars~\citep{Sazonov:2005ri}.

\section{Discussion}
\label{sec:disc}

Our measurement of the broad-band continuum of the GBXE is significantly softer than the spectrum observed by \nustar\ in the Galactic center and than the high-energy \suzaku\ spectrum measured by \citet{Yuasa2012} in nearby regions of the bulge.
There is no detection of the GBXE above the instrumental background level for energies $>$20\,keV.
While \nustar\ measurements of both the diffuse emission and point sources of the inner $\sim$100\,pc indicate a population dominated by IPs with $\langle M_{WD}\rangle \simgt 0.8 M_\odot$, the diffuse bulge spectrum presented here is instead consistent with a large population of DNe, as has been indicated by more recent \suzaku\ studies of the low-energy continuum and Fe line properties of the diffuse emission~\citep{Yamauchi2016,Xu2016,Nobukawa:2017ilu,Koyama:2017zmv} and updated luminosity functions of DNe\citep{Reis:2013bh,Byckling2010}. 

The derived temperature of the bulge, at $kT \approx 8$\,keV, is lower than observed in the central region. 
The \nustar\ spectrum of the diffuse emission in the central 10\,pc is described by a temperature of $kT > 35$\,keV in the energy range above 5\,keV~\citep{Perez2015,Hailey2016}, 
and the combined log$N$-log$S$ population analysis of \chandra\ and \nustar\ point sources in the central 100\,pc implies a $kT \approx$ 20--50\,keV~\citep{Hong2016}. 
Both of these previous \nustar\ studies indicate that the GCXE is dominated by IPs with $\langle M_{WD}\rangle \simgt 0.8 M_\odot$. 
Although a lower GCXE temperature and lighter WD mass were previously implied by low-energy X-ray data~\citep{Muno2004,Muno2009,Heard2013A}, it has been shown that these narrow-band spectral fits bias the derived temperatures to lower values~\citep{Hailey2016}, as they fit only the cooler regions of the accretion flow~\citep{Mukai:2017qww}.

We note that our spectrum is also significantly softer than previous broad-band measurements of the bulge using \suzaku, which warrants discussion.
\citet{Yuasa2012} described the bulge diffuse spectrum in 2--10\,keV and 15--50\,keV using a two-temperature model with $kT_{low} \approx 1.5$\,keV and $kT_{high} \approx 15$\,keV, with the higher temperature having a high intrinsic absorption.
The authors conclude that this emission is primarily due to a combination of IPs with $\langle M_{WD}\rangle = 0.66^{+0.09}_{-0.07} M_\odot$ and soft coronal X-ray sources.

The \nustar\ GBXE measurements presented here are inconsistent with this high-energy \suzaku\ spectrum.
We prepare a simulated 40\,ks observation using the best-fit spectrum from \citet{Yuasa2012}, and combine this with simulated spectra of the 2-bounce GCXE, 0-bounce and 2-bounce CXB, and \nustar\ instrumental background.
Fitting this simulated spectrum to a 1T model yields a $kT \approx 12$\,keV, confirming that the reported \suzaku\ spectrum is significantly harder than the $kT \approx 8$\,keV that we observe. 
Re-fitting our observed spectrum to the same model as \citet{Yuasa2012}, with $kT_{low} \approx 1.5$\,keV and a highly-absorbed $kT_{high}$ that is left free, yields a best-fit $kT_{high} \approx 8$\,keV for all three observations. This is consistent with our derived 1T temperature, though with a poorly-constrained error range of $kT \approx 6-12$\,keV. 
Freezing all temperatures and absorption parameters to those measured by \citet{Yuasa2012} yields a poor fit to the low-energy \nustar\ data, again confirming that our measured spectrum is not consistent with this previous measurement. 

The harder GBXE spectrum observed by \suzaku\ can be explained by differences in the regions measured by \nustar\ and by \citet{Yuasa2012}.
The regions in \citet{Yuasa2012}, while covering out to radii of $r \approx 3^\circ$, also extend to $r <1^\circ$ of the center. 
This inner region is much closer to the inner $\sim100$\,pc of the Galactic center, where \nustar\ measured a hard IP spectrum. 
The \suzaku\ measurements thus cover a transition region, with their spectrum consisting of a combination of the softer DNe spectrum of the bulge and the harder IP spectrum of the center. 

To allow direct comparison of our continuum spectral shape with previous broad-band measurements of the bulge and ridge that were interpreted as arising from a dominant IP population, we have also modeled our GBXE spectrum using an IP mass model.
Our bulge spectrum is consistent with that measured in the ridge by \integral~\citep{Krivonos2007} and \rxte~\citep{Revnivtsev2009}, which both used broad-band measurements to derive $\langle M_{WD}\rangle \approx 0.5 M_\odot$. 
If interpreted as arising from a dominant population of IPs, our GBXE measurement thus implies a much lighter average mass of WDs in these systems than in the central region. 

However, we argue that this derived WD mass is not correct, as the underlying population is not dominated by IPs. 
Our GBXE temperature is not consistent with \suzaku\ measurements of isolated IPs, which showed an average temperature of $kT = 34.0 \pm 4.54$\,keV~\citep{Xu2016}. 
The WD mass is also significantly smaller than the average WD mass in CV systems measured by SDSS, $\langle M_{WD}\rangle = 0.83 \pm 0.24 M_\odot$~\citep{Zorotovic2011}.
Instead, the bulge temperature is consistent with the average temperature of isolated DNe, $kT= 10.7 \pm 2.04$\,keV for an X-ray selected sample~\citep{Xu2016} or $kT = 8^{+10}_{-3}$\,keV for an optically selected sample~\citep{Reis:2013bh}. 
This makes sense, as the inner accretion flow of a non-magnetic CV is cooler than the standoff shock and accretion column of a mCV. 

This conclusion supports the recent \suzaku\ studies of the Fe line emission properties and low-energy continuum of the GDXE. 
\citet{Yamauchi2016}, \citet{Xu2016}, \citet{Nobukawa:2017ilu}, and \citet{Koyama:2017zmv} have shown that the Fe-K$\alpha$, Fe-He$\alpha$, and Fe-Ly$\alpha$ lines have scale heights and equivalent widths that vary between the GCXE, GBXE, and GRXE, indicating different populations in each region. 
\citet{Yamauchi2016} notes that the EWs indicate that a small fraction ($\sim$10-20\%) of the bulge and ridge emission is due to mCVs, and that the majority are non-magnetic CVs, ABs, and coronally-active stars. 
\citet{Xu2016} further argued that the temperature and Fe EWs of the bulge indicate that the majority of sources are DNe.
In addition, \citet{Nobukawa:2017ilu} showed that the low-energy continuum of the bulge is described by a soft bremsstrahlung with $kT \approx 5.1$\,keV, which is inconsistent with a hard IP population.

Recent measurments of the X-ray luminosity function of DNe also support that they could be a dominant contributor to the Galactic diffuse emission. 
\citet{Reis:2013bh} and \citet{Byckling2010} have shown that there is a population of DNe in the solar neighborhood with luminosities down to at least $L_{3-20\,keV} = 10^{30}$\,erg\,s$^{-1}$, with spectra similar to the $kT \approx 8$\,keV observed in brighter DNe.
This could indicate a high space density of short-period, low-luminosity CVs~\citep{2012MNRAS.419.1442P}, as predicted by population synthesis models~\citep{deKool1992,Politano1996}.
However, the statistics of these faint DNe population studies are still poor.

We thus conclude that while the diffuse hard X-ray emission of the Galactic center is dominated by IPs, the emission of the Galactic bulge is dominated by DNe. 
As we state above, our spectrum of the bulge is consistent with previous large-scale measurements along the Galactic ridge; however, there is evidence that the emission along the ridge may be more complicated than a single dominant population. 
\nustar\ measurements of resolved point sources in the Norma arm region ($l \approx 336-339^{\circ}$) revealed an average temperature of $kT \approx 15$\,keV~\citep{Fornasini2017}, significantly harder than the average ridge spectrum measured by \integral\ and \rxte~\citep{Krivonos2007,Revnivtsev2009}. 
This could indicate that the ratios of different source populations in each region are influenced by its particular star formation history, or that the sources resolved by \nustar\ are biased towards harder spectra. 
Follow-up studies of the broad-band spectra of individual sources and 0-bounce studies of the diffuse spectra in the Norma region may help resolve this issue. 

\section{Conclusion}
\label{sec:conc}

We have used a technique that combines contributions from both the \nustar\ focused (``2-bounce") and unfocused (``0-bounce") fields-of-view to measure the broad-band continuum spectrum of the diffuse bulge X-ray emission. 
This allows us to accurately measure the temperature of the low-energy continuum and search for any possible high-energy population, as well as compare with previous broad-band measurements of the inner  $\sim$10\,pc and inner $\sim$100\,pc of the Galactic center using the same instrument.

We detect the GBXE above the instrumental background up to energies of 20\,keV. 
Our measured emissivity, $L_{3-20\,keV}/M \approx (3.4 \pm 0.3) \times 10^{27}$\,erg\,s$^{-1}$\,$M_\odot^{-1}$ is consistent with that measured by \rxte\ in the bulge and ridge.
No significant emission is detected above 20\,keV, and thus we do not find any evidence of an IP population that has been posited to dominate the hard X-ray emission.
The spectrum, which is well-described by a 1T model with $kT \approx 8$\,keV, is significantly softer than the \nustar\ measurements of the GCXE,  where the point source and diffuse populations are consistent with an average temperature of $kT > 20$\,keV~\citep{Hong2016, Perez2015}.
This is also softer than the broad-band \suzaku\ measurements of the GBXE reported in \citet{Yuasa2012}. However, these measurements extended to radii $r<1^\circ$ of the Galactic center, and thus cover a transition region where the spectrum consists of a combination of the softer emission of the bulge and the harder emission of the center.
We thus conclude that previous broad-band X-ray measurements of the bulge were likely contaminated by regions containing hard X-ray emission from the heavy intermediate polar population closer to the Galactic center.
The broad-band continuum of the bulge is consistent with a dominant population of DNe, supporting the same conclusion reached using detailed \suzaku\ studies of the Fe line properties and low-energy continuum of the bulge and updated luminosity distribution measurements of local DNe.
Future observations are necessary to further characterize the transition region between the dominant IP population of the center and this DNe population of the bulge.

\acknowledgments
We thank Ma\"ica Clavel, Charles J. Hailey, Jaesub Hong, Kaya Mori, and John Tomsick for helpful comments and discussions.
K.P. receives support from the Alfred P. Sloan Foundation. 
R.K. acknowledges support from the Russian Science Foundation (grant 19-12-00369).
D.R.W. acknowledges support for this work from NASA Astrophysics Data Analysis grant 80NSSC18K0686 and the \nustar\ mission.

\bibliography{NuSTAR_GBXE}

\begin{appendices}

\section{Validation of 0-bounce, 2-bounce, and background treatment using Obs 60001050005}
\label{app:cxb}

We validate our treatment of the 0-bounce, 2-bounce, and instrumental background emission components by measuring the cosmic X-ray background (CXB) spectral shape and flux. 
For this study, we use Obs 60001050005, an extragalactic field with the point-like source NGC4051, and apply similar data cleaning as described in Sec.~\ref{sec:obs}. 
We remove a 60\arcsec-radius region around this source.  
We also remove data from passages through the South Atlantic Anomaly and any periods with an elevated 3--5\,keV count rate. 
We then extract spectra from the full remaining FOV in two configurations: occulted data, when the \nustar\ FOV is blocked by the Earth, and science-mode data, which is typically used for analysis.

\begin{figure*}
   \begin{centering}
   \subfloat{%
        \includegraphics[width = 0.47\linewidth]{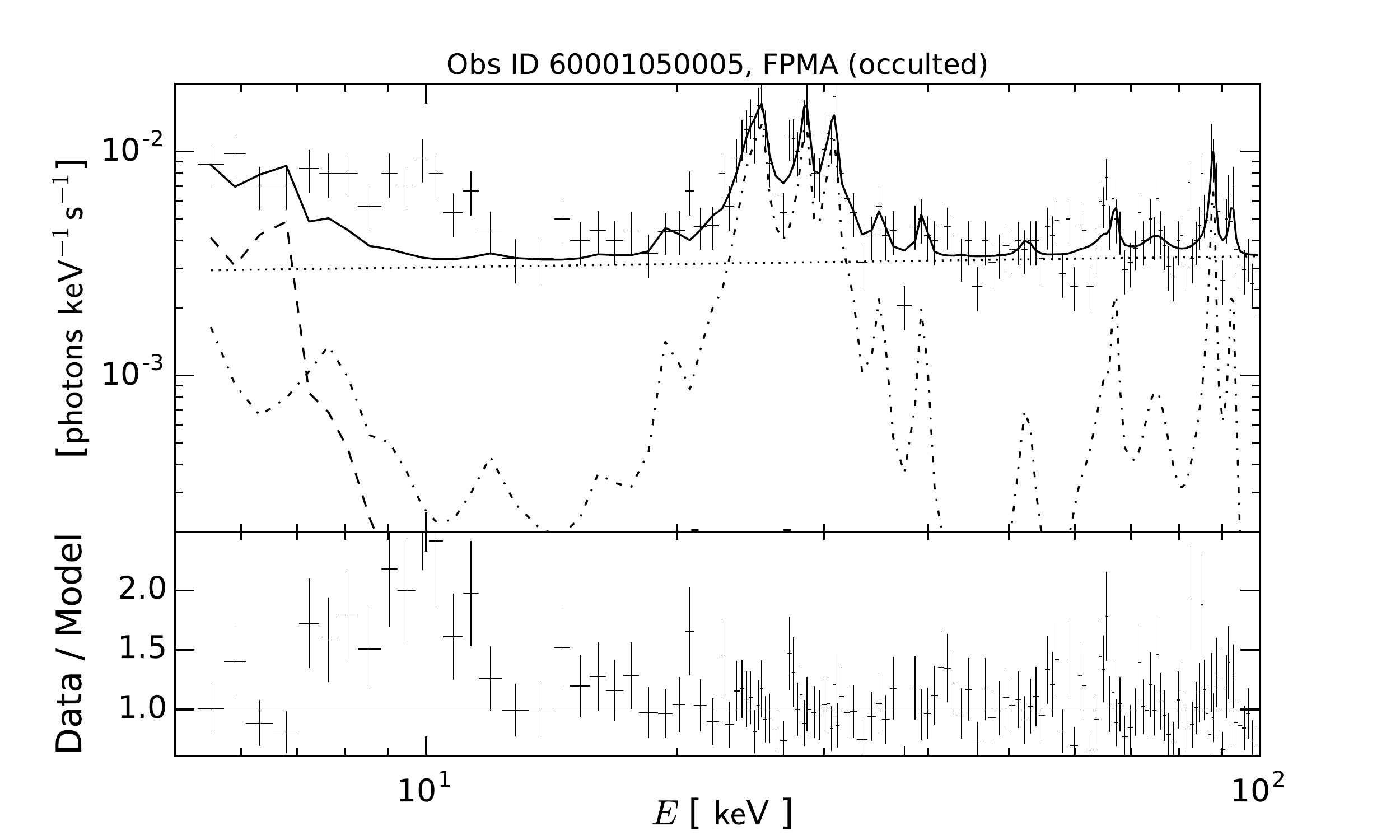}               
      }\qquad
   \subfloat{%
        \includegraphics[width = 0.47\linewidth]{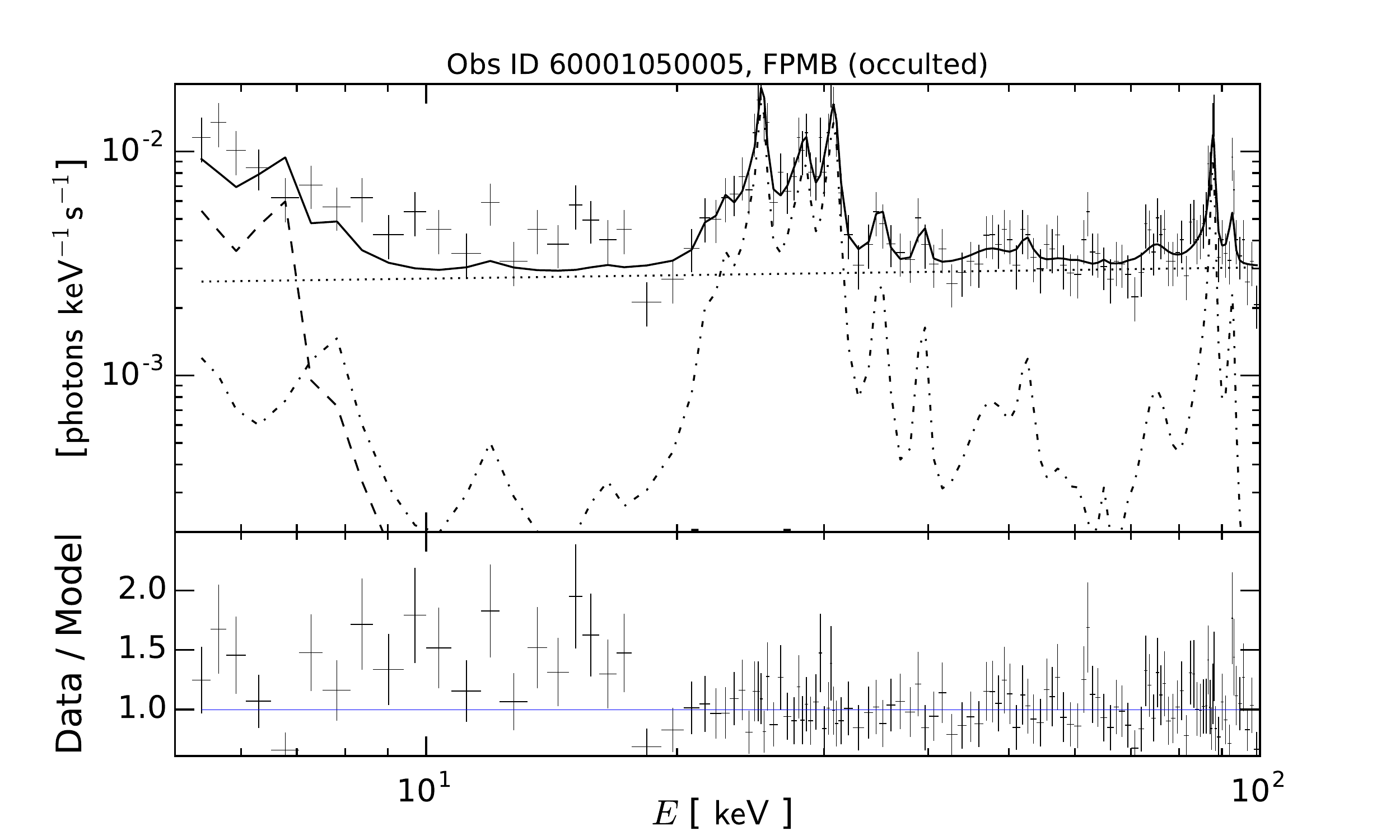}               
      }\qquad
   \subfloat{%
        \includegraphics[width = 0.47\linewidth]{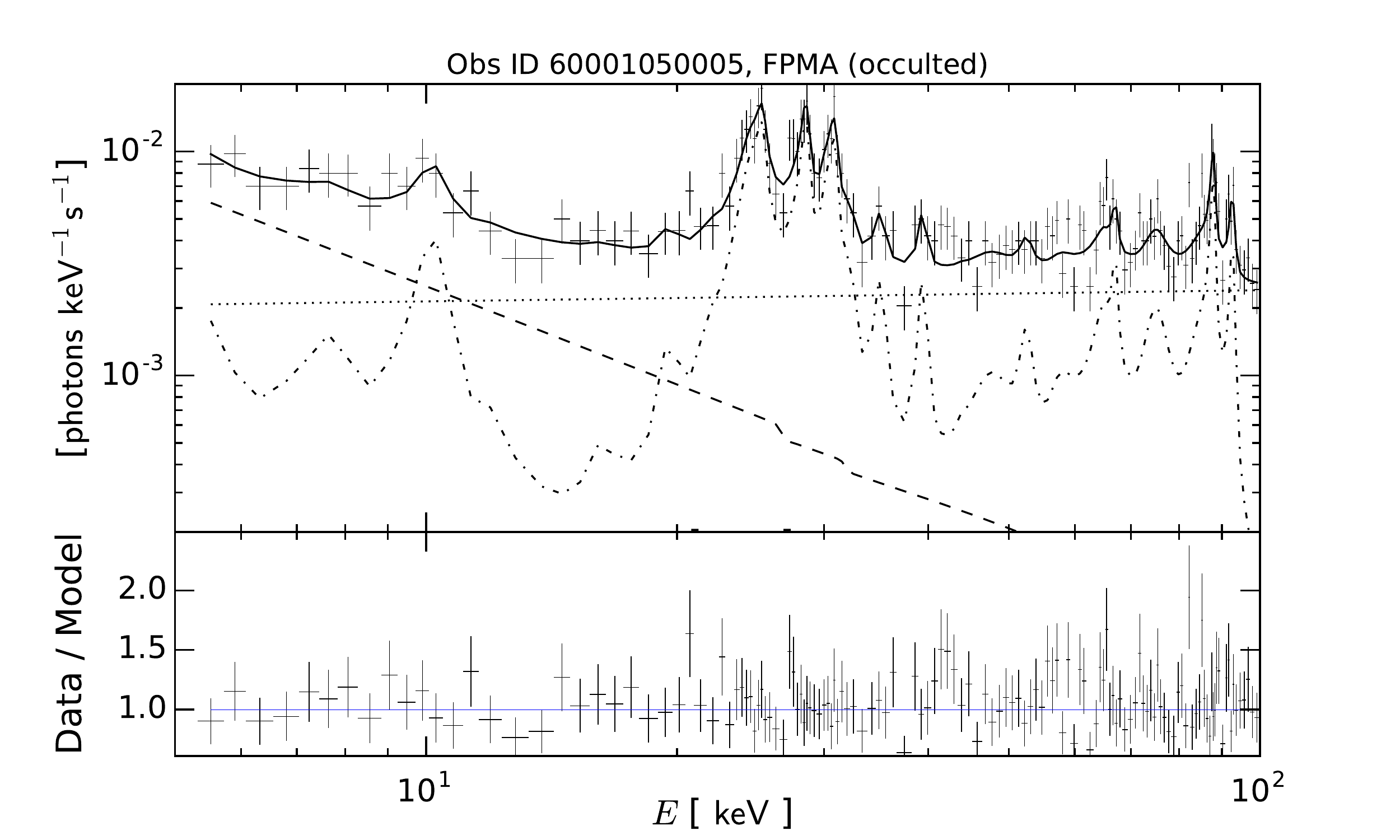}               
      }\qquad
   \subfloat{%
        \includegraphics[width = 0.47\linewidth]{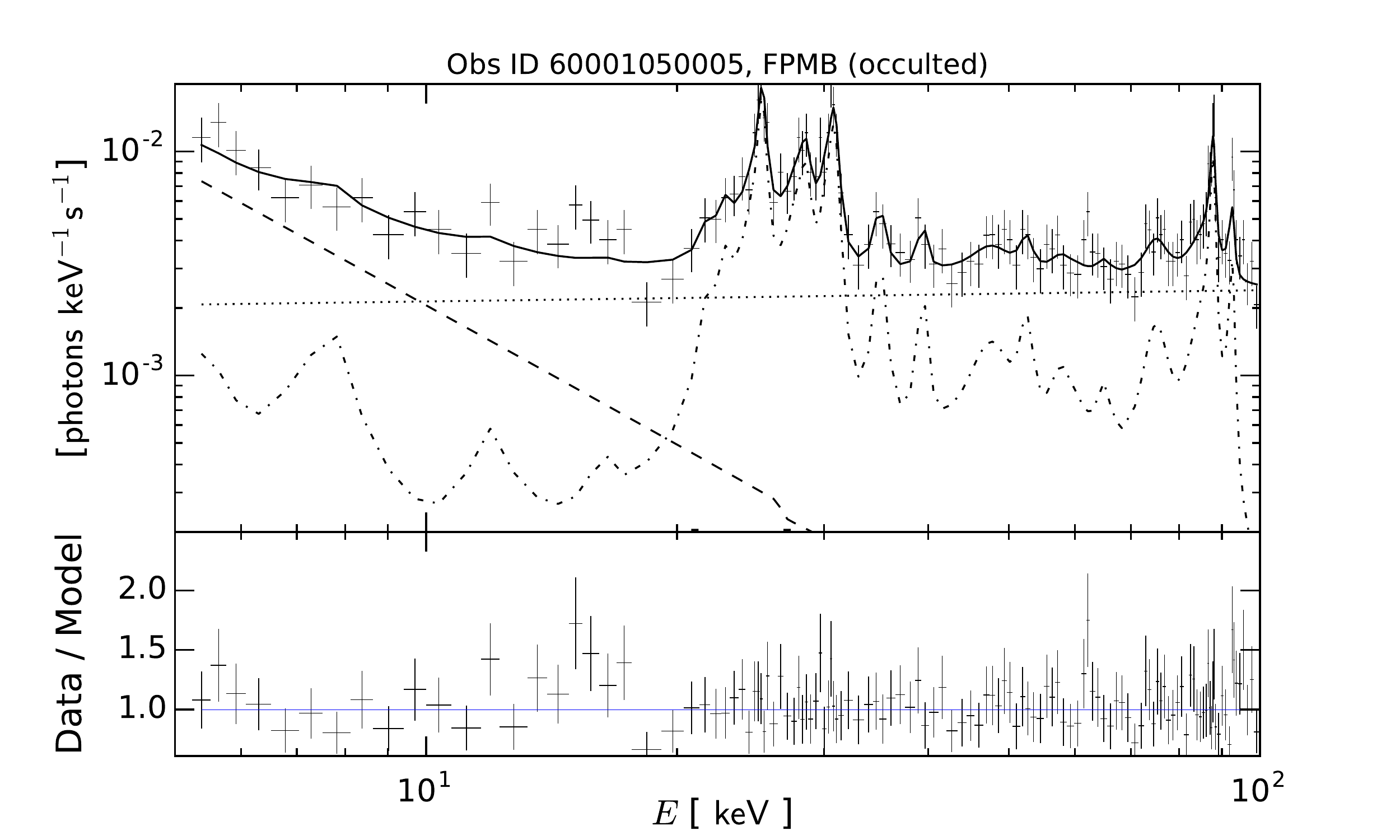}               
     }\qquad
    \caption{\label{fig:occ} Occulted data from FPMA (left) and FPMB (right) of the extragalactic observation 60001050005, with the source NGC4051 removed. The spectra are fit to the default \nustar\ background model (top) and our modified background model (bottom), in which the $\sim$1\,keV thermal plasma component is replaced with a power-law. The model components are illustrated separately, with line emission shown by the dash-dot line, continuum emission shown by the dotted line, and the $\sim$1\,keV thermal plasma or power-law component shown by the dashed line. The modified background model better describes the $E < 20$\,keV spectra. 
}
     \end{centering}
\end{figure*}

We use the occulted data to investigate our instrumental background models, as shown in Fig.~\ref{fig:occ}.
The two background models that we compare are introduced in Sec.~\ref{sec:bkgd}. 
The default \nustar\ background model~\citep{Wik:2014boa} exhibits significant residual emission below 20\,keV. 
In contrast, our modified background model, in which the $\sim$1\,keV thermal plasma component is replaced with a power-law, removes any significant features from the fit residuals, yielding a good fit quality with $\chi^2/d.o.f. =$ 0.95 (107.91/114) and 0.84 (84.63/101) for FPMA and FPMB, respectively.

We then apply our modified instrumental background model to the science-mode data. 
We fix the background power-law index and relative normalization with respect to the high-energy continuum to the best-fit values from occulted data, while the line and continuum component normalizations are left free.
The 0-bounce and 2-bounce effective areas are treated as described in Sec.~\ref{sec:specprep}.
We force the flux of the 2-bounce component, in units of $\rm erg\,s^{-1}\,cm^{-2}\,deg^{-2}\,keV^{-1}$, to be equal to that of the 0-bounce component, since we do not expect significant variations in the flux between the two regions.
The CXB spectral shape frozen to that measured by \integral~\citep{Churazov:2006bk}, with the flux of 0-bounce CXB left free to fit. 

\begin{figure*}
   \begin{centering}
   \subfloat{%
        \includegraphics[width = 0.47\linewidth]{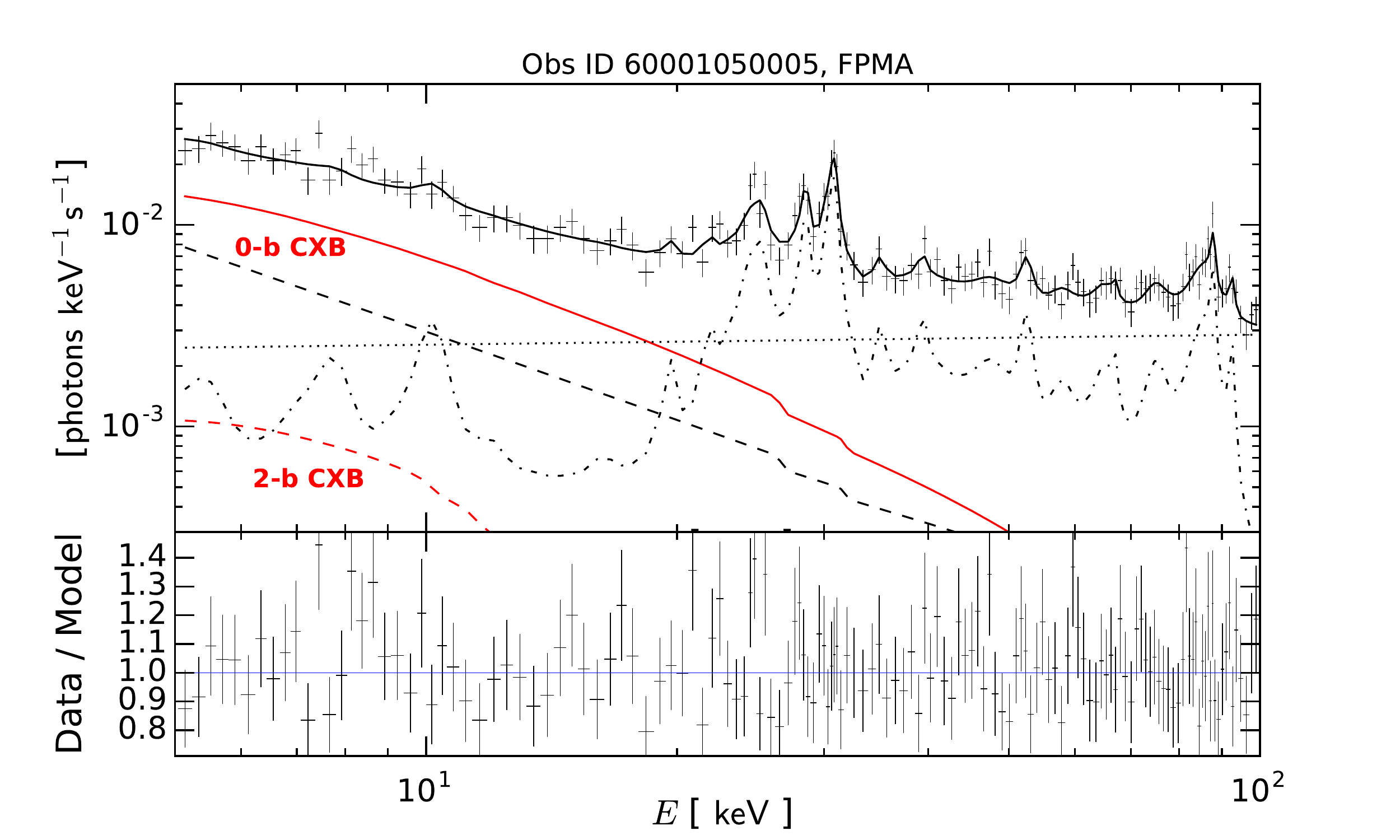}               
      }\qquad
   \subfloat{%
        \includegraphics[width = 0.47\linewidth]{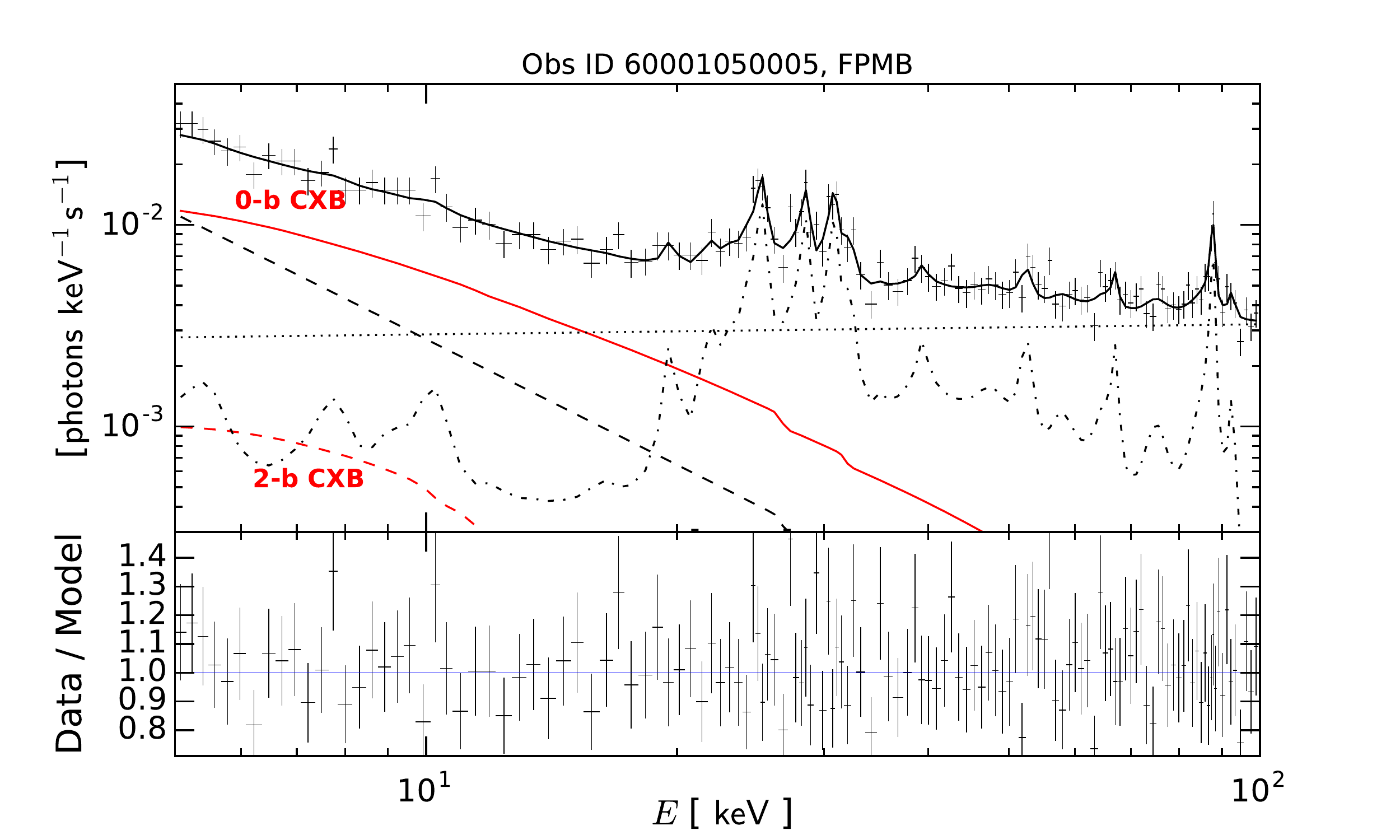}               
     }\qquad
    \caption{\label{fig:cxb} Science-mode data from FPMA (left) and FPMB (right) of the extragalactic observation 60001050005, with the source NGC4051 removed. The spectra are fit to a model consisting of 0-bounce CXB emission (red, solid), 2-bounce CXB emission (red, dash), and instrument background. We use our modified instrument background model, with line (dash-dot), continuum (dot), and power-law (dash) emission. The CXB spectral shape is frozen to that from \citet{Gruber:1999yr,Churazov:2006bk}. Our observed CXB flux agrees with these previous measurements. 
}
     \end{centering}
\end{figure*}

This treatment of the 0-bounce, 2-bounce, and instrumental emission provides a good description of the data, as shown in Fig,~\ref{fig:cxb}. 
The overall fit quality is good, with $\chi^2/d.o.f. =$ 0.88 (105.14/120) and 0.87 (94.06/108) for FPMA and FPMB, respectively. 
The derived 3-20\,keV CXB flux is $F = (3.6 \pm 0.5) \times 10^{-11} \rm erg\,s^{-1}\,cm^{-2}\,deg^{-2}\,keV^{-1}$ for FPMA and $F = (3.3^{+0.6}_{-0.5}) \times 10^{-11} \rm erg\,s^{-1}\,cm^{-2}\,deg^{-2}\,keV^{-1}$ for FPMB. 
This is consistent with the 3--20\,keV flux measured by \integral, $F = (2.6 \pm 0.5) \times 10^{-11} \rm erg\,s^{-1}\,cm^{-2}\,deg^{-2}\,keV^{-1}$, given the $\sim$10\% overall \nustar\ cross-calibration uncertainty~\citep{Madsen:2015jea}. 
We thus conclude that our analysis method yields the correct overall spectral shape and flux level for the CXB.

\end{appendices}

\end{document}